\documentclass[%
 reprint,
 floatfix,
 amsmath,amssymb,
 aps,
 prl
]{revtex4-2}

\usepackage{graphicx}
\usepackage{dcolumn}
\usepackage{bm}

\usepackage{siunitx}
\usepackage{tablefootnote}

\begin{document}

\preprint{APS/123-QED}

\title{Impulse-driven transport of liquid metal from z-pinch electrodes and implications for fusion fuel contamination}


\author{Daniel P. Weber}
\author{Colin S. Adams}
    \email{csadams@vt.edu}
\affiliation{Virginia Polytechnic Institute and State University}

\date{\today}
             
\begin{abstract}
    The Liquid Electrode eXperiment is a surrogate environment to study the dynamic behavior of liquid metal plasma facing components relevant to a z-pinch device. Current pulses with amplitudes between \num{50} and \num{200}~\unit{\kilo\ampere} produce magnetic fields up to \num{30}~\unit{\tesla} at the surface of a wire \num{1.5}--\num{2.5}~\unit{\milli\meter} in radius, mounted with one end submerged in a pool of liquid metal. The resulting forces generate a fast-moving annular jet surrounding the wire, preceded in some cases by small ejected droplets of varying sizes. High-speed videography records the motion of the liquid metal free surface upon exposure to magnetic pressures between \num{0.5} and \num{10}~\unit{\mega\pascal}. The vertical velocity of the resulting jets ranged from \num{0.6}--\num{5.3}~\unit{\meter/\second} with consistent radial expansion. The velocities of the ejecta ranged from \num{-3.1} to +\num{18.9}~\unit{\meter/\second} in the vertical direction and from \num{-14.3} to +\num{6.3}~\unit{\meter/\second} in the radial direction. We investigate the likelihood and severity of z-pinch core contamination for repetitively-pulsed plasmas in the context of the observed droplets for liquid metal plasma facing component candidate materials: lithium, gallium, silver, tin, lead, FLiBe, and FLiNaK.
\end{abstract}

\keywords{Suggested keywords}

\maketitle


\section{Introduction}
Liquid metal (LM) plasma-facing components (PFCs) may be preferable to solid surfaces in fusion reactors due to several beneficial characteristics: liquids exhibit self-healing behavior, are capable of mitigating extreme heat loading, and may be used to breed tritium if the LM contains lithium. As most fusion concepts employ hydrogenic fuels, previous experiments and simulations have examined the effects plasmas have on LMs \cite{deuterium_addition_LiSn,trit_breed_impact,new_facility_Flili}, as well as the extent of adsorption of these fuels into the LMs and methods to recycle such material \cite{hyd_retent_implications_high_z,Gas_recycling_bubbling}. Previous research efforts have examined the effect of the plasma on the LM leaving questions about how the LM might affect the fuel during confinement. A relevant body of research focuses on the production of vapor shields from LMs. Elements such as lithium and tin have been shown to produce vapor clouds over LM surfaces, which can limit the heat flux to PFCs \cite{vapor_shielding,Li_divertor_design,Sn_Vapor_Shielding}. These clouds have shown to exist relatively close to the LM, but diffuse material contaminating the core fuel remains a concern.

Magneto-hydrodynamics induced fluid movement is expected in many devices which employ LM PFCs, yet there have been few investigations of these dynamics. An experiment in the early 2000s \cite{small_MHD_waves} excited small amplitude surface waves in a pool of liquid gallium with an applied background magnetic field ($B_0 = \num{50}~\unit{\milli\tesla}$). MHD-like damping of the wave amplitude was demonstrated in the presence of the field. MHD effects were also studied in another investigation \cite{free-surface_MHD_Heat} seeking to use the Lorentz force to control the poloidal LM flow in a tokamak for the purposes of heat removal. This study considered bulk surface movement as well as the potential for small-scale instability that could lead to core contamination. When a PFC is used as an electrode, such as in a z-pinch, the LM will be subject to strong Lorentz forces beyond those previously studied. There remain open questions regarding the response of LMs to large, localized currents and magnetic fields.

Due to the extreme environment intended for the candidate plasma facing materials, reactor concept studies have sought to identify materials that are compatible and functional in neutron-rich environments. Pure lithium blankets have been considered for tritium breeding \cite{nstx_lld_li_pfc,Li_HIDRA,vapor_shielding}, but mixtures with other elements are also often considered, such as lead-lithium or tin-lithium \cite{LiSn_divertor,lisn_pfc_behavior,critical_exploration}. One beneficial property of heavier additives is the ability to slow and multiply free neutrons in order to improve the probability of tritium breeding. Beryllium, zirconium, and lead are considered ideal neutron multipliers for this purpose\cite{neutron_multipliers}. Alloys of alkali salts are particularly suited to tritium breeding and there is an established institutional knowledge base due to their importance in nuclear fission. The alloys LiF-BeF$_2$ (FLiBe) and LiF-NaF-KF (FLiNaK) have low melting points, contain lithium, and can multiply neutrons. Similarly, tin has been shown to improve the overall heat load removal \cite{LiSn_divertor,Sn_Vapor_Shielding} of liquid lithium while not increasing the hydrogenic species adsorption \cite{deuterium_addition_LiSn}. Finally, gallium and its alloys are considered for their very low melting point and low vapor pressure at high temperatures \cite{sn_cps_iter,Ga_Li_Divertor}.

We report results from an investigation of the MHD response of LMs to localized, z-pinch scale currents and magnetic fields, including liquid behavior which has not yet been examined with the relevant geometry or current densities. A statistical analysis of droplet trajectories quantifies the likelihood of fusion core contamination by ejected droplets. As this is a topic of concern for future z-pinch fusion devices, contamination limits are estimated for various LM PFC candidates, above which a z-pinch core is expected to enter radiative collapse.

\section{Methods}

\begin{figure}[ht]
    \centering
    \includegraphics[width=\linewidth]{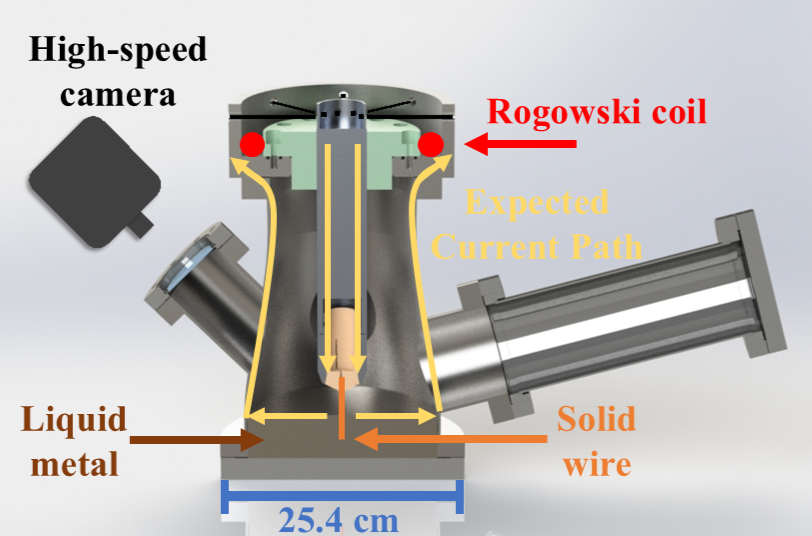}
    \caption{LEX apparatus showing approximate positions of the diagnostic instruments and key features. All instruments are insulated from the chamber to protect from from the current discharge. The solid wire is interchangeable for radii variation and replacement of damaged wires.}
    \label{fig:exp_apparatus}
\end{figure}

In the Liquid Electrode eXperiment (LEX), a solid wire acts as an electromagnetic surrogate for a stable z-pinch column, simplifying the interpretation of experimental data by eliminating challenges associated with plasma-material interactions and long-time column stabilization. The coaxial design, shown in Figure \ref{fig:exp_apparatus}, injects current into the center of the liquid electrode where large Lorentz forces perturb the otherwise equilibrium surface.

A pulse forming network (PFN) transmits up to \num{186}~\unit{\kilo\ampere} to the pool of LM through a purpose-designed coaxial feedthrough and load wires of varying radii, submerged with one end below the LM surface to produce current densities at the LM surface comparable to those near a z-pinch. The current conducts radially away from the wire on the LM surface and the self-generated azimuthal magnetic fields produce locally strong Lorentz forces on the LM, which range \num{0.5}--\num{10}~\unit{\mega \pascal}. This range is accessible by adjusting both the charging potential of the PFN and the wire diameter. Since magnetic pressure decays proportional to $1/r^2$ radially and exponentially in the axial direction, there are comparably strong radial and axial gradients. In LEX, the LM pool is a eutectic tin-bismuth alloy heated to a nominal temperature of \num{433}~\unit{\kelvin}, which exceeds the melting point of \num{411}~\unit{\kelvin}.

Spatial and temporal data are collected using two key diagnostics. A Rogowski coil placed around the inner electrode measures the change in magnetic flux \cite{hutch} which is integrated to infer current through the wire. The surface deflection is imaged by a high speed camera at a nominal rate of 20,000 frames per second (fps) and oriented at a \num{45}\unit{\degree} angle to the liquid surface. The camera is calibrated for metrology using an evenly spaced checkerboard grid and a bi-linear interpolation scheme maps pixels to positions in the calibration plane.

\begin{figure*}[h]
    \centering
    \includegraphics[width=\linewidth]{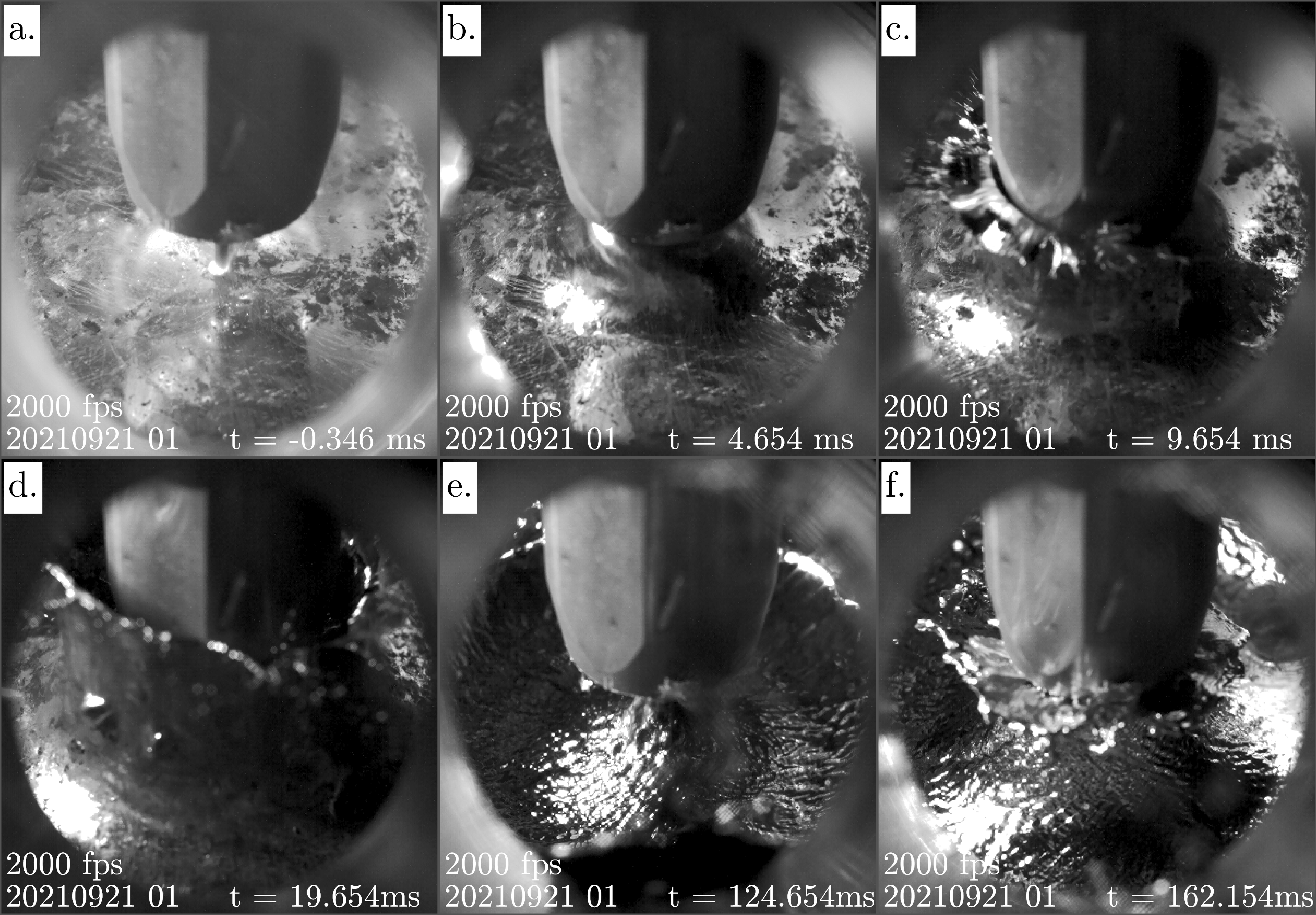}
    \caption{The response of the LM during (a.) and after (b.-f.) the discharge of the current pulse. This qualitative behavior is consistent in all pulsed trials reported here, regardless of driving current or wire diameters. This specific sequence is from an applied current $\approx$\num{50}~\unit{\kilo \ampere} through a \num{1.59}~\unit{\milli \meter} radius wire.}
    \label{fig:long_time}
\end{figure*}

As the current discharges, an annular depression forms around the wire with a diameter a few times that of the wire. After the pulse, small ejecta droplets are seen in most trials travelling upward from the edge of the depression with both positive and negative radial velocities. Concurrently, the depression appears to expand axially and radially below the LM surface but observations are limited to the first few milliseconds. An annular LM jet rises from around the wire and travels vertically for some time, obscuring optical access to the wire-liquid interface. The evolution of the jet is shown in Figure \ref{fig:long_time}. In cases with the most drive energy, material from this jet reaches the height of the vacuum containment vessel after colliding with the collet that holds the wire. After approximately \num{100}~\unit{\milli \second}, the jet structure breaks apart and a large cavity becomes visible.

\section{Annular Liquid Metal Jet}

\begin{figure*}
    \centering
    \includegraphics[width=\linewidth]{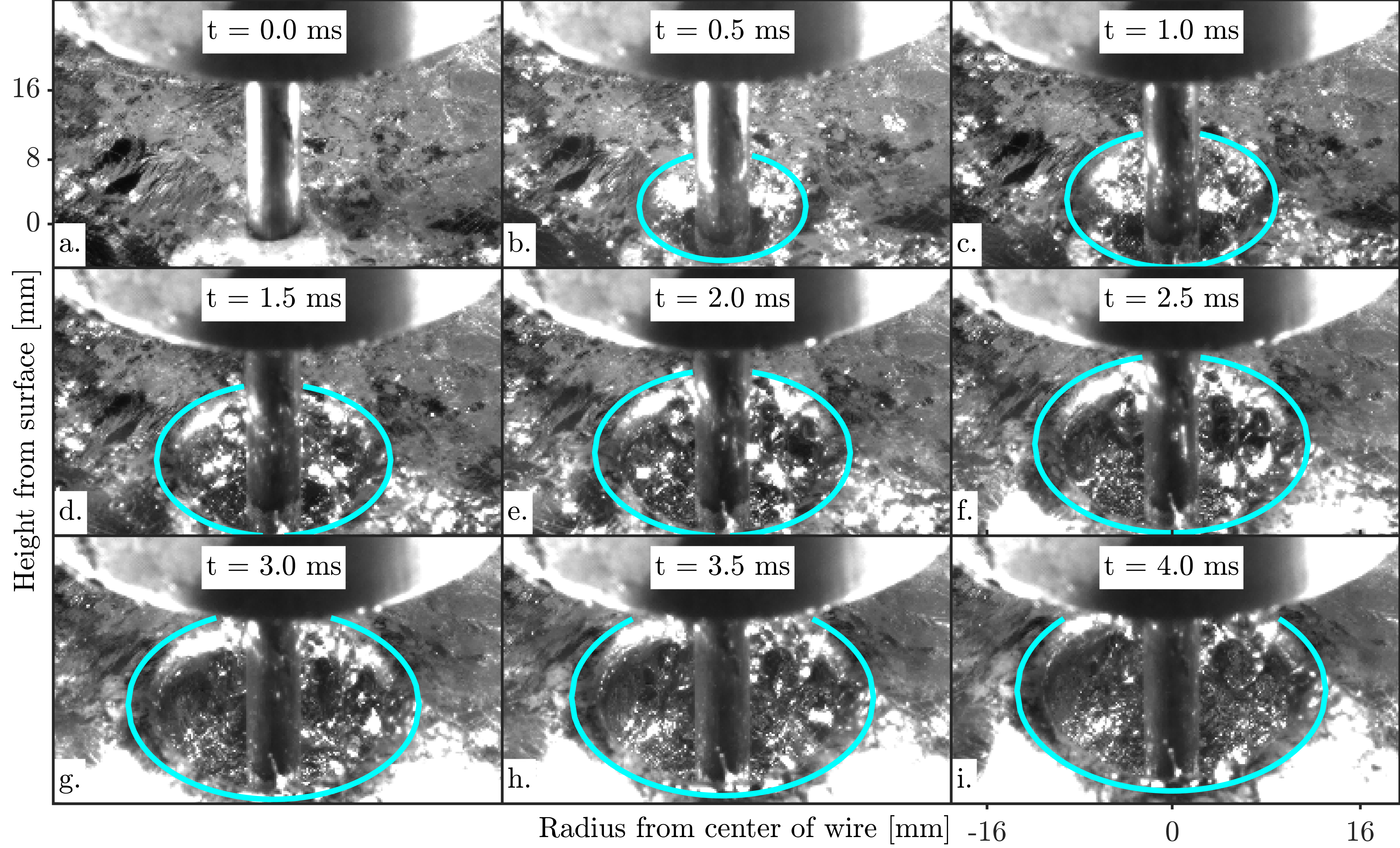}
    \caption{Ellipses fit to the annular jet crest to estimate the radius and height which are used to infer velocity. Images are captured with a \num{50}~\unit{\milli\meter} focal length lens at a rate of \num{20}~\unit{kfps}, every tenth image is shown here.}
    \label{fig:ellipse_show}
\end{figure*}

Ellipses are fit to the apparent crest of the annular jet every tenth frame using the horizontal extrema and corresponding vertical position at the center of the wire, shown in Figure \ref{fig:ellipse_show}. The jet radius and vertical height are estimated from the bi-linear interpolation of the horizontal extrema to ensure that the measurement is in the calibrated $r$-$z$ plane and therefore breaks the spatial ambiguity from a single camera view. Time of flight of features is employed to infer the radial and vertical velocity. For the frames recorded before the jet reaches the collet, there is very little change in the vertical velocity, so an average is calculated for each trial.

The reported estimates of magnetic pressure ($p_{\mathrm{m}}$) result from approximating the surface magnetic field as that produced by an infinitely long wire, accounting for exponential decay in current below the LM surface due to the skin effect. Accordingly, the magnetic pressure
\begin{equation}\label{eq:mag_pres}
    p_{\mathrm{m}} = \frac{B_{\theta}^2}{2\mu} = \frac{\mu I_{\mathrm{z,rms}}^2}{8\pi^2 r^2}\exp{\left(-\frac{2d}{\delta}\right)},
\end{equation}
where the magnetic permeability, $\mu$, is the weighted sum of the respective permeabilities based on the weight percentage (42\% tin, 58\% bismuth) and $r$ is the radius from the wire. Surface elevation of the liquid, $z=z_0$, is referenced relative to the bottom of the pool, $z=0$ so the depth below the surface can be considered $d = z_0-z$. Skin depth,

\begin{equation}
    \delta = \sqrt{\frac{2\eta}{\omega\mu}},
\end{equation}
is calculated from the e-folding depth for TEM waves in a conductor, assuming low frequencies. The resistivity of the LM, $\eta$, is also a weighted sum of the individual resistivites of tin and bismuth. A Fourier transform of the current pulse indicates a driving frequency of $f =  \num{11.9}~\unit{\kilo\hertz}$.

Since the current pulse is underdamped, the net electromagnetic force over the timescales relevant to the LM response is approximated by a representative square pulse. The amplitude of the representative pulse is chosen to be the root-mean-square (RMS) of the inferred current,
\begin{equation}
    I_{\mathrm{rms}} = \sqrt{\frac{1}{N}\sum\limits_{i=1}^{N}(I_i^2)},
\end{equation}
where $N$ is the number of data points collected during the current discharge and $I_i$ is the current measurement at a specific data point. The duration of the square pulse is the length of time of current fluctuations, $t_{\mathrm{pulse}}(N) = \num{265}~\unit{\micro\second}$.

\begin{figure}
    \centering
    \includegraphics[width=\linewidth]{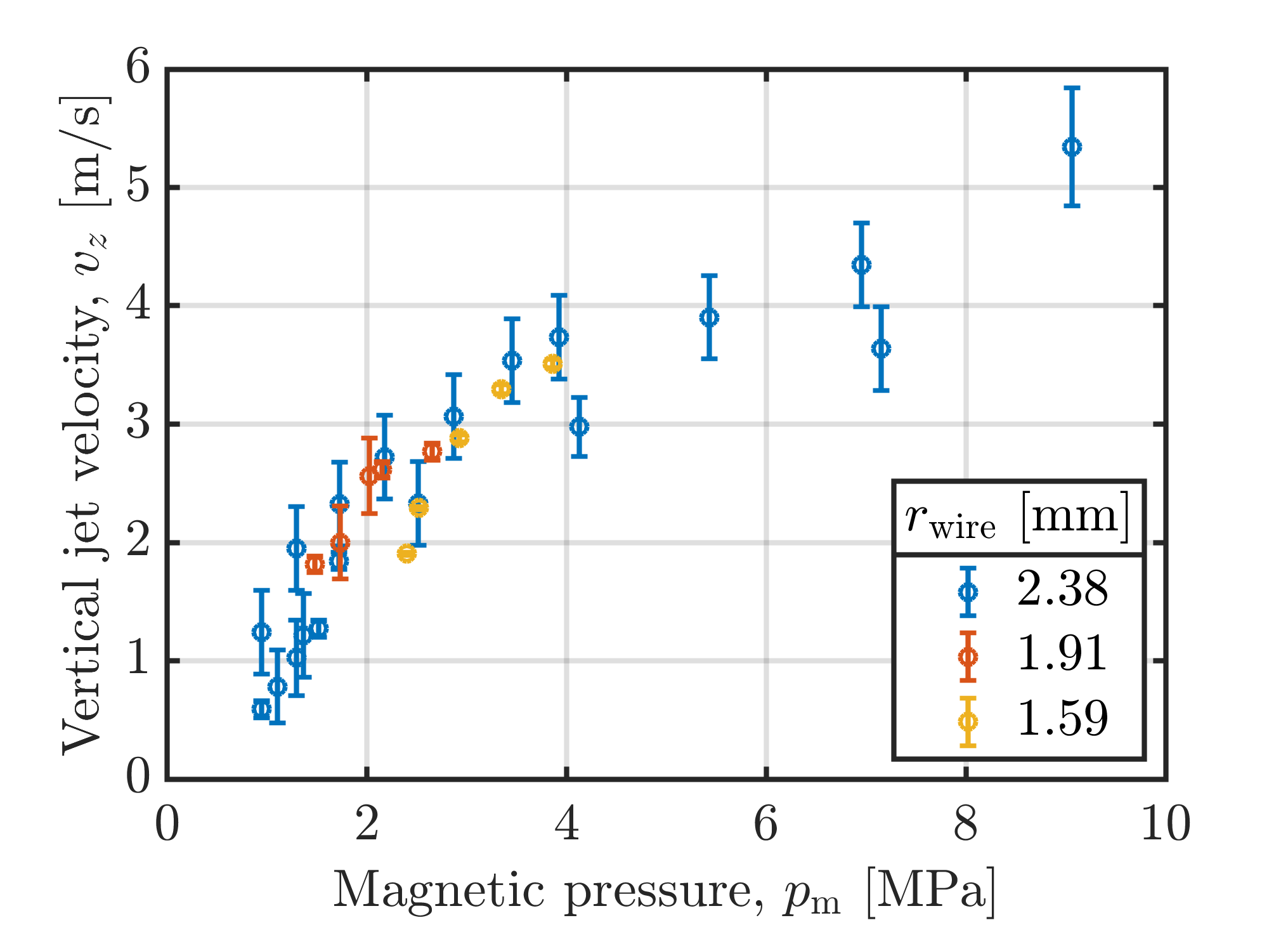}
    \caption{Annular jet vertical velocity ($v_z$) in response to the applied magnetic pressure ($p_{\mathrm{m}}$). The pressure is varied by changing the wire radii ($r_{\mathrm{wire}}$) and the applied PFN potential.}
    \label{fig:vertvel_vs_pml}
\end{figure}

After incrementally increasing the applied magnetic pressure on the LM surface, a direct correlation to the vertical velocity of the annular jet is observed, as shown in Figure \ref{fig:vertvel_vs_pml}. In other scenarios where spray is produced by driving pressures, liquid velocities have a square-root dependence on the driving pressures \cite{odd}, which is likely the greatest contributor to the apparent trend of the measurements taken over magnetic pressure space. Across the range of magnetic pressures investigated, the response of the LM is qualitatively similar over tens of milliseconds, as represented in Figure \ref{fig:long_time}.

In repetitively-pulsed devices featuring LM PFCs, the boundary geometry for subsequent punches may depend on the evolution of this annular jet. Large-scale fluid motion is observed to settle around \num{250}~\unit{\milli\second}, which could limit operations to a few Hertz, less than the \num{10}~\unit{\hertz} proposed rate for commercial z-pinch reactors \cite{zap_approach}.

\begin{figure*}[ht]
    \centering
    \includegraphics[width=\linewidth]{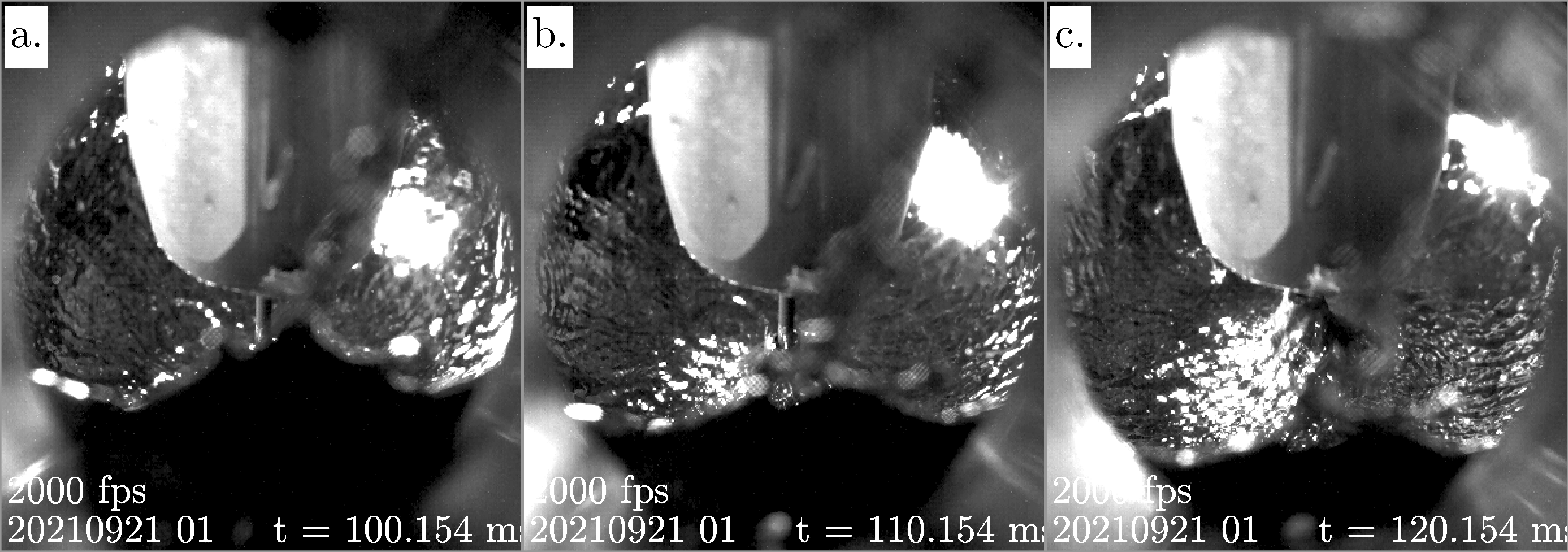}
    \caption{During the dynamic response of the LM, a nearly flat surface re-forms around \num{100}~\unit{\milli\second} after the current pulse, which could be an ideal time for a subsequent pinch in a repetitively pulsed device (a.). This morphology lasts for approximately \num{10}~\unit{\milli\second} until a conical recoil begins to travel up the length of the wire (b.). This electrode feature could help to reduce the downward Lorentz force on the LM surface and extends \num{47}~\unit{\milli\meter} to the height of the collet after another \num{10}~\unit{\milli\second}, until (c.).}
    \label{fig:ideal_time}
\end{figure*}

To improve the repetition rate, an ideal time during the bulk LM movement for a subsequent pulse in this experiment is identified as between \num{100}--\num{110}~\unit{\milli\second}, shown in Figure \ref{fig:ideal_time}. The LM exhibits a concave, hemispherical free-surface at this time, nearly flat below the wire which provides a surface close to the equilibrium position. A follow-on pulse during this time is anticipated to produce similar LM behavior to the first pulse. Between \num{110} and \num{120}~\unit{\milli\second} is also identified as a potential frame for the second pulse where a conical-shaped fluid surface forms in the center, as shown in Figure \ref{fig:ideal_time}b-c. In this scenario, current flowing at the liquid surface would now posses an axial component, rather than being purely radial. In turn, the axial component of the Lorentz force would be reduced over a substantial portion of the liquid surface, possibly resulting in smaller amplitude metal response.

\section{Ejected Droplets}\label{sec:ejecta}

\begin{figure}
    \centering
    \includegraphics[width=\linewidth]{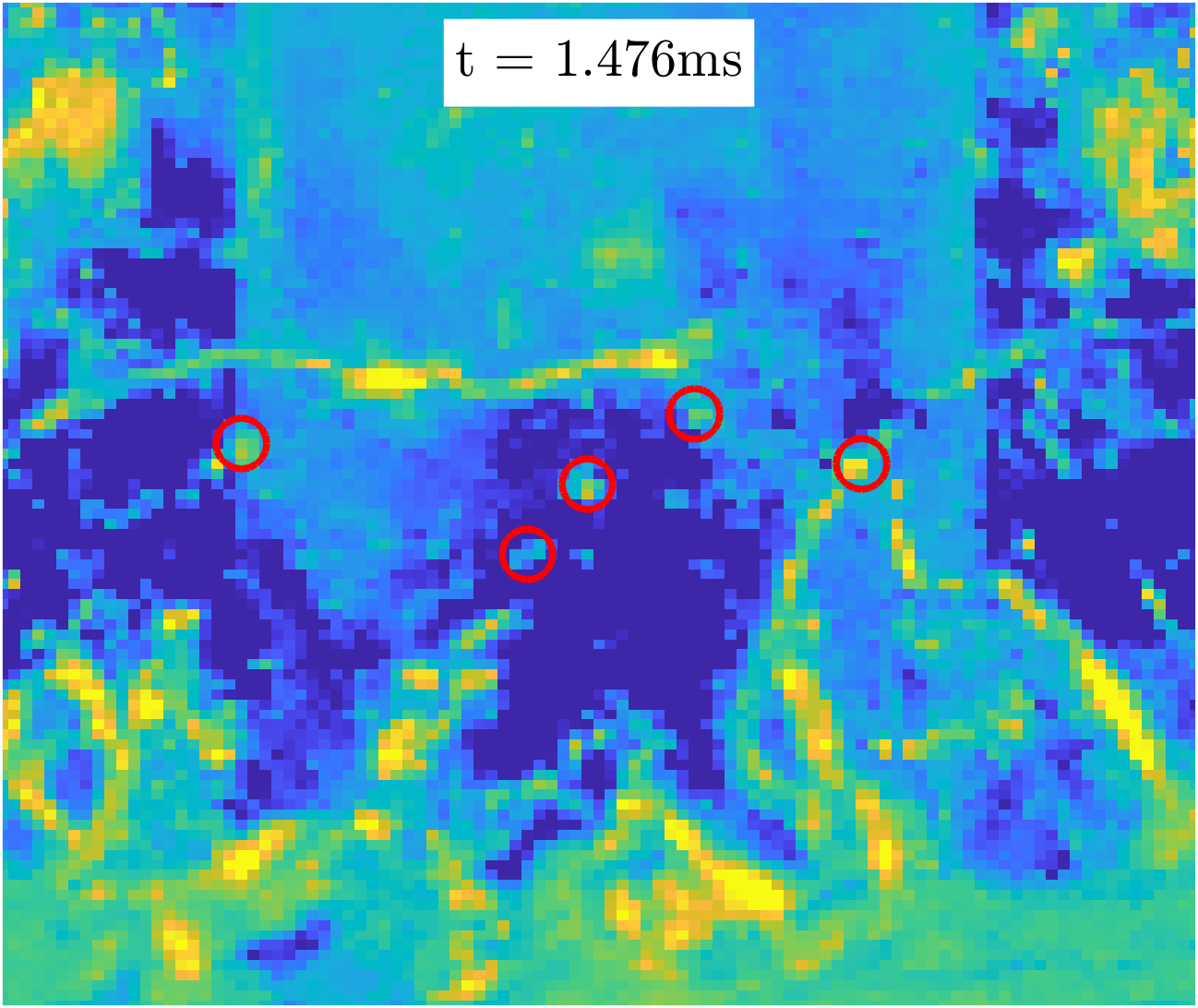}
    \caption{Wire-centered closeup image of the LM response with circles around centroids of identified ejecta droplets. Image taken with a \num{130}~\unit{\milli\meter} focal length lens at \num{20}~\unit{kfps}. Features displayed using false color and intensity follows a logarithmic scaling.}
    \label{fig:ejecta_ident}
\end{figure}

Immediately following the current pulse, small ejecta, roughly \num{0.3}~\unit{\milli\meter} in diameter are observed moving above the surface of the LM, as shown in Figure \ref{fig:ejecta_ident}. These droplets are traced for 6 sequential frames in order to estimate their trajectory and average velocity before passing out of the field of view. The diameter of each droplet is estimated to be the average of the vertical and horizontal diameters to account for non-spherical shape. The droplet is tracked in each frame using the brightest point within its perimeter. The two lights used to illuminate the inside of the chamber are $\pi/2$ and $-\pi/2$ offset from the camera, respectively, in the azimuthal direction. The bidirectional perpendicular illumination should make the relative center of approximately spherical droplets appear brightest. The estimated droplet diameters are averaged through time and mass of the droplet is estimated by assuming a spherical volume with the density of tin-bismuth. Ejecta radial and vertical velocities are inferred by tracking the motion of the brightest point from frame to frame, assuming no azimuthal velocity, and reporting the average over time.

Often, the only resolvable droplets are located in front of the wire and out of the calibration plane, as shown in Figure \ref{fig:ejecta_ident}. It is difficult to ensure droplets that are not located in front of the wire are in the plane of calibration. As a result of recording the LM dynamics from a single camera perspective, droplets in front of the calibration plane have a larger apparent size and a reduction in apparent radial velocity. Motion toward or away from the camera is estimated based on the rate of change of the size of droplets across frames. This approximation of the out-of-plane velocity is recorded and indexed in time for droplets that start near the calibration plane.

The positions of the droplets were used to estimate the radial velocity. Corrections to apparent radial velocity ($\tilde{v_r}$) were evaluated by breaking the volume into two distinct regions where droplets are tracked.
$$
    v_r = \frac{\tilde{v_r}}{\cos(\theta)},
$$
The region determines the approximate azimuthal angle, $\theta$, which the droplet makes from the calibration plane with a liberal error estimate. The region where the droplets appeared in front of the wire (Region 1) is where $\theta = \pi/3 \pm \pi/9$, and the regions on either side of the wire (Region 2) are where $\theta = \pi/6 \pm \pi/6$.

 \begin{figure}[h]
    \centering
    \includegraphics[width=\linewidth]{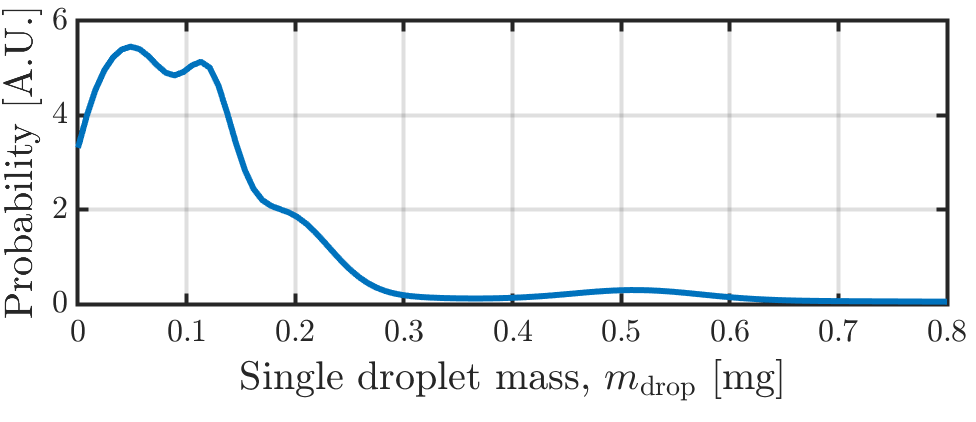}
    \caption{A finite mixture probability distribution based on the estimated ejecta mass, assuming spherical ejecta droplets. Estimates are from tests with a wire radius of \num{1.59}~\unit{\milli\meter} and mean magnetic pressures of \num{3}~\unit{\mega\pascal}. The contributing Gaussian distributions are truncated at zero to eliminate non-physical probabilities.}
    \label{fig:mass_prob}
\end{figure}

To determine the mass distribution, estimated masses of droplets with resolvable radii are used to calculate a finite mixture probability density function. A normal distribution, $p_i(x)$, represents each estimated time-averaged droplet mass. The width of the distribution represents the larger of either the standard deviation of the individual droplet mass estimate or the propagated error. Propagation of error accumulates from sources of uncertainty including the droplet diameter, position, angle with the calibration plane, and measurements from a single camera. A sum of the normal distributions with weights, $w_i$ such that $\sum\limits_i w_i = 1$, produces a probability density function, $f(x)$, which can be updated with data from successive measurements,

\begin{equation}
    f(x) = \sum\limits_{i=1}^n w_ip_i(x).
\end{equation}

This approach is a useful way to model a distribution function based on discrete measurements without assuming a particular functional form for the overall distribution. Because the droplet formation mechanism is still unknown, the weights for each estimation were equal when generating the mixture model. The resulting finite mixture model of estimated droplet mass is shown in Figure \ref{fig:mass_prob}. This mass model predicts most probable droplet sizes based on previous responses from the subset of tests of a wire radius of \num{1.59}~\unit{\milli\meter} and an applied magnetic pressure of \num{3}~\unit{\mega\pascal}. For these conditions, the most probable droplet mass is \num{48.85}~\unit{\micro\gram}. However, the resolution of the camera limits measurements to ejecta diameters greater than $limit$, establishing an artificial cutoff where there were no $p_i(x)$ with probable mass below $limit2$. Therefore, the reduction in probability as $m\rightarrow0$ may be artificial, but the probability density function does still lend insight into the probability of large mass droplets.

\begin{figure}[ht]
    \centering
    \includegraphics[width=\linewidth]{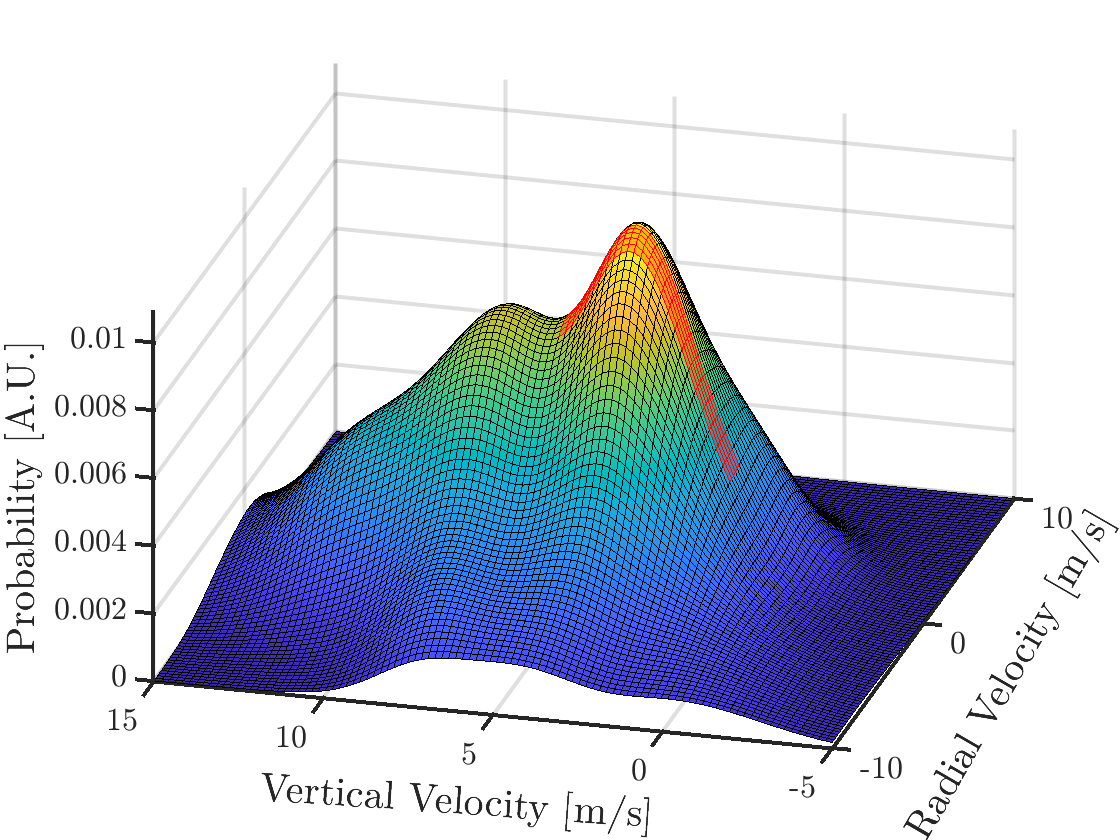}
    \caption{The bi-variate finite mixture probability density function for ejecta velocities based on experimental measurements within LEX from tests with a wire radius of \num{1.59}~\unit{\milli\meter} and mean magnetic pressures of \num{3}~\unit{\mega\pascal}. Integration between two radial and vertical velocity limits determine the likelihood of a tin-bismuth droplet forming and intersecting the core plasma during a subsequent pinch, shown as red overlain section.}
    \label{fig:vert/radvel_ejecta}
\end{figure}

Measured radial and vertical velocities of the ejecta are used to estimate the velocity probability space using the same finite mixture model, assuming velocity in the azimuthal direction ($v_{\theta}$) is zero. For each droplet, inferred radial and vertical velocities are treated as quantities that depend upon each other. Therefore, a bi-variate normal distribution represents the average radial and vertical velocities, simultaneously. The correlation between the two velocity components is described by calculating the covariance matrix ($\sigma_{ij}$) of the distributions which represents the two-dimensional width of each of the contributing normal distributions. The covariance is calculated using the square-exponential kernel
\begin{equation}
    \sigma_{ij} = \sigma_i\sigma_j\exp\left(-\frac{(x_i-x_j)^T*(x_i-x_j)}{2l^2}\right).
\end{equation}
The finite mixture method uses equal weighted bi-variate distributions to predict the likelihood of ejecta formed with combinations of radial and vertical velocities, shown in Figure \ref{fig:vert/radvel_ejecta}. As with the mass model, the velocity model uses velocity estimates from tests with a wire radius of \num{1.59}~\unit{\milli\meter} and average applied magnetic pressures of \num{3}~\unit{\mega\pascal}.

In a repetitively-pulsed fusion device, the trajectories of droplets formed during a pulse could cause them to move through regions where a pinch is formed during the subsequent pulse. To quantify the probability of contamination of the fusion fuel by these droplets, we examine droplets with radial and vertical velocities that risk placing a droplet in the subsequent pinch region. An adiabatic scaling of recent z-pinch experimental results is used by \citet{shumlak_z_pinch} to forecast pinch geometry at $Q=1$, where
\begin{equation}
    Q = \frac{P_{\mathrm{fusion}}}{P_{\mathrm{input}}},
\end{equation}
which results in predicted pinch radii of \num{0.6}~\unit{\milli\meter}. The assembly region of the apparatus discussed by Shumlak is \num{500}~\unit{\milli\meter} in length and has a radius of \num{100}~\unit{\milli\meter}. In an energy-producing reactor, pinches with these approximate dimensions would be repetitively formed many times every second.

Should droplets ejected during each pulse translate into the region where the next pinch is about to be formed, there is a potential that these droplets could contaminate the fuel, increasing radiative energy losses. Accordingly, radial velocity limits are developed based on the maximum or minimum radial distance a droplet would need to traverse in the shortest or longest period of time, respectively. Geometrically, the maximum distance is the sum of the radius of the experiment and the radius of the pinch, thereby assuming a droplet forms at a wall boundary and travels to the opposite edge of the pinch region. The minimum distance is the case of a droplet forming on the electrode within the radial extent of the pinch region, $r =$ \num{0}~\unit{\milli\meter}. The range of possible time of flight of the droplet is estimated from the proposed repetition rate previously discussed of \num{100}--\num{120}~\unit{\milli\second}. "The result of these considerations is that droplets with radial velocity in the range \num{-1}--\num{0}~\unit{\meter/\second} pose a contamination risk to subsequent plasma pulses. Axial velocity limits are similarly based on the geometry of the pinch assembly region and the time scale between pulses. The maximum axial distance is the total length of the pinch region and the minimum is also $z = $\num{0}~\unit{\milli\meter}. Using the same time frame of \num{100}-\num{120}~\unit{\milli\second} previously mentioned, droplets with axial velocity in the range \num{0}--\num{5.5}~\unit{\meter/\second} pose a particularly serious contamination risk.

These limits are applied as limits of integration to the velocity finite mixture probability density function to estimate the probability that a droplet is formed with velocity in the range that poses contamination risk. The integration region is illustrated in Figure \ref{fig:vert/radvel_ejecta} as the red shaded area over the probability density function. According to this model, there is a 4.5\% chance that a droplet forms with a trajectory that would carry it to the pinch-forming volume in the time between pulses. Although this estimation is applicable to Sn-Bi, a wire radius of \num{1.59}~\unit{\milli\meter}, and a magnetic pressure of \num{3}~\unit{\mega\pascal}, the approach can be extended to develop insight regarding the consequences of LM--plasma interactions in future fusion devices.

\section{Implications for Other Alloys}\label{sec:cand_alloy}

The droplet sizes and velocities observed for the liquid tin-bismuth may not be representative of the droplets ejected by other metals; the material properties of other liquid PFCs will likely differ significantly from those in the experiments described here. In order to estimate the distribution of droplet size and velocity for other LMs, we develop a model to estimate these distributions by comparing the viscosity and surface tension of other materials to tin-bismuth. This section treats velocities and masses as dependent and independent random variables, respectively. Accordingly, the implications regarding the potential impacts of other plasma-facing materials are based on the empirical statistics reported in the previous section for tin-bismuth.

Other materials were chosen that could be relevant to LM PFCs with both low vapor pressures and low melting points. Droplet properties are estimated for lithium (Li), gallium (Ga), silver (Ag), tin (Sn), lead (Pb), LiF-BeF$_2$ (FLiBe), and LiF-NaF-KF (FLiNaK), as shown in Table \ref{tab:cont_prob_cand}. Lithium is a candidate plasma-facing material for fusion devices due to the prospect of tritium breeding. Silver and lead nuclei have large cross sections for the neutron multiplying (n,2n) reaction which benefits the breeding of tritium from lithium. FLiBe and FliNaK are included in this study for reference; while these are not clear candidates for liquid electrodes, these mixtures have frequently been considered  for use in scenarios where high current densities may be present during magnetic disruptions in tokamaks.

Consider the scenario that the same magnetic impulse is applied to a static liquid, as in the experiments in LEX described previously. A similar current and magnetic pulse creates the same force density $\vec{f}$. Simplifying the Cauchy momentum equation for conditions dominated by viscosity and momentum to describe the droplet movement results in
        
\begin{equation}\label{eq:cauchy}
    \rho\frac{D\vec{v}}{Dt} = \vec{f} = \mu\hat{\nabla}^2\vec{v}.
\end{equation}
Gravitational forces are assumed to be negligible due to the relatively short time scales. Furthermore, the fluid that separates into a droplet must be moving faster than the rest of the body, which indicates the fluid pressure has a limited effect on the droplet formation and can also be ignored. We assume that the pressure forces set the fluid in motion initially, but significantly decrease in magnitude by the time the droplets are formed.

For constant force density in Equation \ref{eq:cauchy}, the viscosity ($\mu$) has an inverse relationship with the velocity Laplacian, $\hat{\nabla}^2\vec{v}$, which is proportional to the acceleration of the fluid element. Since both viscosity and density ($\rho$) are inversely proportional to acceleration, the acceleration of droplets of other materials ($\vec{a}_Z$) is approximated by scaling the acceleration of the tin-bismuth droplets by the ratio of their viscosities and densities and assuming zero initial velocity
        
\begin{equation}
    \vec{a}_{Z} = \frac{\mu_{\mathrm{Sn-Bi}}}{\mu_{Z}}\frac{\rho_{\mathrm{Sn-Bi}}}{\rho_{Z}}\vec{a}_{\mathrm{Sn-Bi}}.
\end{equation}
Since we expect droplet velocities to be proportional to acceleration for a given current discharge, we compare the velocities of ejecta from plasma-facing liquids by multiplying the velocities measured in tin-bismuth by the same ratios. Velocity distributions for these candidate liquids are calculated with the same multivariate method previously discussed. The probability of contamination for each material is estimated by integrating the respective probability density function over velocity space as discussed previously.


While droplet acceleration is estimated from viscosity and density, the droplet mass itself may change for liquids of differing surface tension and acceleration. In order to relate the surface tension of liquids to the size and mass of the droplet ejected, we turn to the pendant drop test \cite{pendant_drop} which is commonly used to evaluate liquid surface tension. This test involves a syringe of liquid suspended above a plate. Liquid is forced out of the syringe which forms a droplet that is initially attached to the bottom of the needle through capillary forces, which is a function of the surface tension and the area of the needle's opening. As more liquid is extruded, the droplet eventually reaches a mass where the force of gravity exceeds the retaining capillary force. In the resulting force balance, the acceleration $\vec{a}$ of the largest possible liquid drop extruded from a tube of diameter $d$ can be related to the maximum droplet mass, $m$, and the surface tension of the liquid $\gamma$ via Equation \ref{eq:pendant_drop}.

\begin{equation}\label{eq:pendant_drop}
    m\vec{a} = \gamma \pi d.
\end{equation}
Because the mass is proportional to the surface tension, the mass of a droplet from each plasma-facing material is approximated by scaling the most probable mass from tin-bismuth by the ratio of the surface tension for the two liquids

\begin{equation}
    m_{Z} = \frac{\gamma_{Z}}{\gamma_{\mathrm{Sn-Bi}}}\frac{m_{\mathrm{Sn-Bi}}}{\vec{a}_{Z}}.
\end{equation}

All referenced viscosities are calculated from an equation of either empirical or theoretical fit as a function of temperature of the form

\begin{equation}
    \log\left[\mu_Z(T)\right] = -A + \frac{B}{T}.
\end{equation}
The densities and surface tensions are calculated from empirical or theoretical equations of the form

\begin{equation}
    \rho(T) = C - D\cdot (T - T_{\mathrm{ref}}),\
\end{equation}
\begin{equation}
    \gamma(T) = E - F\cdot (T - T_{\mathrm{ref}}).
\end{equation}
In these equations, the coefficients $A$, $B$, $C$, $D$, $E$, and $F$ are found from sources in literature included references, included in the respective columns of Table \ref{tab:cont_prob_cand}.

The viscosity, density, and surface tension of tin-bismuth (Sn-Bi) are interpolated to the temperature of the LM pool in LEX, $T_{\mathrm{LEX}} = \num{423}~\unit{\kelvin}$, from data reported in literature \cite{gamma_mu_snbi}. The most probable mass of tin-bismuth used in these approximations is the peak of the mass distribution shown in Figure \ref{fig:mass_prob}, which is \num{48.9}~\unit{\micro\gram}.

\begin{table*}[]
    \centering
    \begin{tabular}{|c|c|c|c|c|c|c|c|}
        \hline
        Candidate, & $T_{\mathrm{opt}}$  & $\mu_{Z}(T_{\mathrm{opt}})$ & $\rho_{Z}(T_{\mathrm{opt}})$ & $\gamma_{Z}(T_{\mathrm{opt}})$ & $n_{\mathrm{vap}}(T_{\mathrm{opt}})$ & $P_{\mathrm{drop}}$ & $m_{\mathrm{ejecta}}$\\
        $Z$ & [K] & $\overline{\mu_{\mathrm{Sn-Bi}}(T_{\mathrm{LEX}})}$ & $\overline{\rho_{\mathrm{Sn-Bi}}(T_{\mathrm{LEX}})}$ & $\overline{\gamma_{\mathrm{Sn-Bi}}(T_{\mathrm{LEX}})}$ & [m$^{-3}$] & [\%]  & [\unit{\micro\gram}]\\
        \hline
        Li & 800 & 0.13 \hspace{3mm} \cite{mu_li} & 0.06 \hspace{3mm} \cite{mu_li}  & 0.81 \hspace{3mm} \cite{mu_li} & \num{7.89e24} \hspace{3mm} \cite{vap_press_metals} & $<$ 0.001 & $<$ 0.001\\
        Ga & 800 & 0.23 \hspace{3mm} \cite{mu_ga} & 0.67 \hspace{3mm} \cite{mu_ga} & 1.50 \hspace{3mm} \cite{gamma_ga} & \num{1.76e13} \hspace{3mm} \cite{vap_press_metals} & 0.05 & 0.252\\
        Ag & 1253  & 1.43 \hspace{3mm} \cite{mu_pb_ag} & 1.08 \hspace{3mm} \cite{mu_pb_ag} & 2.08 \hspace{3mm} \cite{gamma_pure_metals} & \num{3.05e19} \hspace{3mm} \cite{vap_press_metals} & 10.70 & 671\\
        Sn & 1089 & 0.29 \hspace{3mm} \cite{mu_sn} & 0.77 \hspace{3mm} \cite{mu_sn} & 1.19 \hspace{3mm} \cite{gamma_sn} & \num{1.03e16} \hspace{3mm} \cite{vap_press_metals} & 0.13 & 0.648\\
        Pb & 799 & 0.61 \hspace{3mm} \cite{mu_pb_ag} & 1.21 \hspace{3mm} \cite{mu_pb_ag} & 0.98 \hspace{3mm} \cite{gamma_pb} & \num{5.40e17} \hspace{3mm} \cite{vap_press_metals} & 2.47 & 35.1\\
        FLiBe & 860 & 3.23 \hspace{3mm} \cite{mu_flibe} & 0.23 \hspace{3mm} \cite{gamma_mu_flinak} & 0.44 \hspace{3mm} \cite{gamma_mu_flinak} & \num{2.03e14} \hspace{3mm} \cite{flibe_vap_press} & 2.51 & 16.2\\
        FLiNaK & 860 & 1.61 \hspace{3mm} \cite{gamma_mu_flinak} & 0.23 \hspace{3mm} \cite{rho_flinak} & 0.42 \hspace{3mm} \cite{gamma_mu_flinak} & \num{1.98e16} \hspace{3mm} \cite{fluoride_vap_press} & 0.48 & 1.52\\
        \hline
    \end{tabular}
    \caption{Thermo-physical properties for certain LM PFC candidates at respective optimal temperature ($T_{\mathrm{opt}}$), vapor density at the optimal temperature ($n_{\mathrm{vap}}(T_{\mathrm{opt}})$), contamination probability of a single droplet ($P_{\mathrm{drop}}$), and approximate mass of all droplet contamination ($m_{\mathrm{con}}$). Ratios in reference to tin-bismuth are interpolated from data in \citet{gamma_mu_snbi} to the temperature in LEX ($T_{\mathrm{LEX}}$). Candidate material specific references are included in table.}
    \label{tab:cont_prob_cand}
\end{table*}

To determine the best operating temperature for each candidate alloy or mixture, an optimization scheme was developed with the objective of reducing the total estimated mass ($m_{\mathrm{tot}}$) from both ejecta droplets ($m_{\mathrm{ejecta}}$) and the static vapor ($m_{\mathrm{vap}}$),

\begin{equation}
    m_{\mathrm{tot}} = m_{\mathrm{ejecta}} + m_{\mathrm{vap}}.
\end{equation}

A more comprehensive picture of probable contamination mass was estimated from the mass of a single droplet ($m_{\mathrm{drop}}$). Assuming there is a linear correlation to the probability of single droplet contamination ($P_{\mathrm{drop}}$) and the estimated number of droplets generated per pulse from a tin-bismuth pool ($N_{\mathrm{ejecta}} = 40$), the single droplet mass is multiplied by both,

\begin{equation}\label{eq:m_ejecta}
    m_{\mathrm{ejecta}} = m_{\mathrm{drop}} P_{\mathrm{drop}} N_{\mathrm{ejecta}}.
\end{equation}
The ejecta mass, therefore, retains the assumption the number of droplets per shot remains constant for all candidate liquids. Also retained is the assumption that the pinch volume remains constant, as that was key in developing the statistics for contamination probabilities.

The time between pinches proposed here allows significant diffusion time for a static vapor from the liquid to fill a cylindrical volume ($V_{\mathrm{FuZE}}$) approximating that of presently operational z-pinch devices (cylinder with \num{10}~\unit{\centi\meter} radius and \num{1}~\unit{\meter} length) \cite{zap_approach}. As the pinch is accelerated and assembled, the plasma collects material within the volume as a ``snowplow" effect before forming the pinch column wherein it would collect all gaseous matter which fills the chamber - even contamination. This PFC vapor distribution would not be uniform, but is estimated to be so for simplicity.

\begin{equation}
    m_{\mathrm{vap}} = \left(\frac{p_{\mathrm{vap}}}{k_\mathrm{B}T}\right)\left(\frac{M_Z}{N_{\mathrm{A}}}\right)V_{\mathrm{FuZE}},
\end{equation}
where $M_Z$ is the molar mass of the material and $N_{\mathrm{A}}$ is Avogadro's number. Vapor pressures ($p_{\mathrm{vap}}$) were calculated from best fits of previous data in the form

\begin{equation}
    \log\left[p_{\mathrm{vap}}(T)\right] = G - \frac{H}{T},
\end{equation}
where $G$ and $H$ are coefficients of a least-squares best fit. Thermophysical properties used in this analysis are listed at the optimal temperature of each candidate in Table \ref{tab:cont_prob_cand}.

The estimated probabilities of single-droplet core contamination, shown in Table \ref{tab:cont_prob_cand}, are relatively low (most below 10\%). Because $N_{\mathrm{ejecta}}$ is estimated to be around 40, at least one droplet is expected to reach the core for candidates with a probability $\approx$2.5\% (lead and FLiBe). Multiple droplet contamination is likely for silver, indicated by a single-droplet contamination probability $>$10\%. Table \ref{tab:cont_prob_cand} compares the total forecasted mass of ejecta droplets as modeled by Equation \ref{eq:m_ejecta} for each candidate plasma facing material. The severity of this contamination to the power balance within a fusion plasma depends on the contaminant species.

\section{Severity of Core Contamination}

For each candidate LM, we quantify the potential impact of contamination caused by the presence of droplets near the z pinch core. The experimental tin-bismuth data is utilized to estimate the probability of contamination and the mass of a representative droplets produced from other plasma facing materials. To do this, it is assumed the ejecta is located within the core volume at the time of a subsequent pinch and has been heated to the same temperature of the plasma.

For a pinch in equilibrium, power is balanced by the rates of ohmic heating ($P_{\mathrm{j}}$) and radiative cooling ($P_{\mathrm{r}}$) \cite{high_density_z_pinch}. When the ratio $P_{\mathrm{r}}/P_{\mathrm{j}} > 1$, the pinch contracts and cools into radiative collapse. Alternatively, when $P_{\mathrm{r}}/P_{\mathrm{j}} < 1$, the column expands with the increasing heat.

Considering the addition of contaminants to a core in equilibrium, both energy transfer mechanisms will be affected. Assumptions are first made about the initial conditions of the pinch using the adiabatic scaling \cite{shumlak_z_pinch} of experimental results to forecast necessary conditions at $Q=1$, as mentioned in previously.

A projected plasma temperature of \num{3.5}~\unit{\kilo\eV}, an ion density of \num{2e25}~\unit{\per\meter\cubed}, a pinch radius of \num{0.6}~\unit{\milli\meter}, and pinch current of \num{600}~\unit{\kilo\ampere} are all used for further calculations.

The radiated energy from a core plasma includes components from bremsstrahlung radiation, $P_{\mathrm{Br}}$, and the contaminant radiation, $P_{\mathrm{r,con}}$,

\begin{equation}
    P_{\mathrm{r}} = \int\limits_{V} \left(P_{\mathrm{r,con}} + P_{\mathrm{Br}}\right) \hspace{1mm} dV,
\end{equation}
where bremsstrahlung radiation is estimated assuming an optically thin plasma, which corresponds to the maximum possible power loss,

\begin{equation}
    P_{\mathrm{Br}} = 1.69\times10^{-32}n_eT_e^{1/2} \hspace{3mm}[\mathrm{W/cm^3}].
\end{equation}

A coronal equilibrium model has previously been proposed in literature \cite{coronal_equilibrium} which estimates radiative cooling rates and mean charge states for elements with atomic numbers $2\leq Z\leq 92$ based on bremsstrahlung radiation, radiative recombination, and line transitions, assuming the core plasma is optically thin. Specific rates ($P_{\mathrm{r,con}}$) are estimated for impurities of interest at the core plasma temperature,

\begin{equation}\label{eq:prcon}
    P_{\mathrm{r,con}} = n_e n_Z L_Z \hspace{3mm} [\mathrm{ergs/cm^3/s}].
\end{equation}

In Equation \ref{eq:prcon}, $n_Z$ is the density of the contaminant ions, and $L_Z$ is the loss rate based on the polynomial fits to the model using coefficients $A(i)$,

\begin{equation}
    \log_{10}L_Z = \sum\limits_{i=0}^5 A(i) \left\{ \log_{10}(T_e [\mathrm{keV}]) \right\}^i.
\end{equation}

Ohmic heating ($P_{\mathrm{j}}$) is estimated using the projected current densities,

\begin{equation}
    P_{\mathrm{j}} = \int\limits_{V} \eta_{\mathrm{Sp}} | \vec{J} |^2 dV,
\end{equation}
where the resistivity ($\eta_{\mathrm{Sp}}$) is based on the Spitzer model for resistivity in an optically thin plasma,

\begin{equation}
    \eta_{\mathrm{Sp}} = \frac{4\sqrt{2\pi}}{3}\frac{\langle Z_{\mathrm{eff}}\rangle e^2m_e^{1/2}\ln\Lambda}{(4\pi\varepsilon_0)^2(k_{\mathrm{B}}T_e)^{3/2}}.
    \label{eq:spitz}
\end{equation}

Here, the Coulomb logarithm ($\ln\Lambda$) is calculated based on electron-ion collisions,
\begin{equation}
    \ln\Lambda = 24 - \ln\left(\frac{n_e^{1/2}}{T_e}\right), \hspace{3mm} T_i m_e/m_i < 10Z^2 [\mathrm{eV}] < T_e.
    \label{eq:cou_log}
\end{equation}

In Equation \ref{eq:spitz}, $T_e$ is the electron temperature in the gas where $T_e = T_i = T_{\mathrm{pinch}}/2$. Also, $\langle Z_{\mathrm{eff}}\rangle$ is the effective mean charge state of the possibly contaminated, two species plasma,

\begin{equation}
    \langle Z_{\mathrm{eff}} \rangle = \frac{\langle Z \rangle n_{Z}}{n_{\mathrm{tot}}} + \frac{n_{i,\mathrm{H^+}}}{n_{\mathrm{tot}}},
\end{equation}
where  $n_{i,\mathrm{H^+}}$ is the hydrogen ion density, $n_{\mathrm{tot}} = n_{i,\mathrm{H^+}} + n_{Z}$ is the total plasma density, and $\langle Z\rangle$ is the mean charge state of the contamination based on the polynomial fits to the model using coefficients $B(i)$,

\begin{equation}
    \langle Z \rangle = \sum\limits_{i=0}^5 B(i)\left( \log_{10}T_e[\mathrm{keV}] \right)^i.
\end{equation}

The coronal equilibrium model did not estimate mean charge states nor radiation rates for gallium and lead \cite{coronal_equilibrium}. Linear fits from the nearest neighbors that were included (zinc and arsenic for gallium and mercury and bismuth for lead) are used to approximate these values. Though the increase is not linear, the small step size between elements lends this to be a viable approximation.

\begin{figure}[ht]
    \centering
    \includegraphics[width=\linewidth]{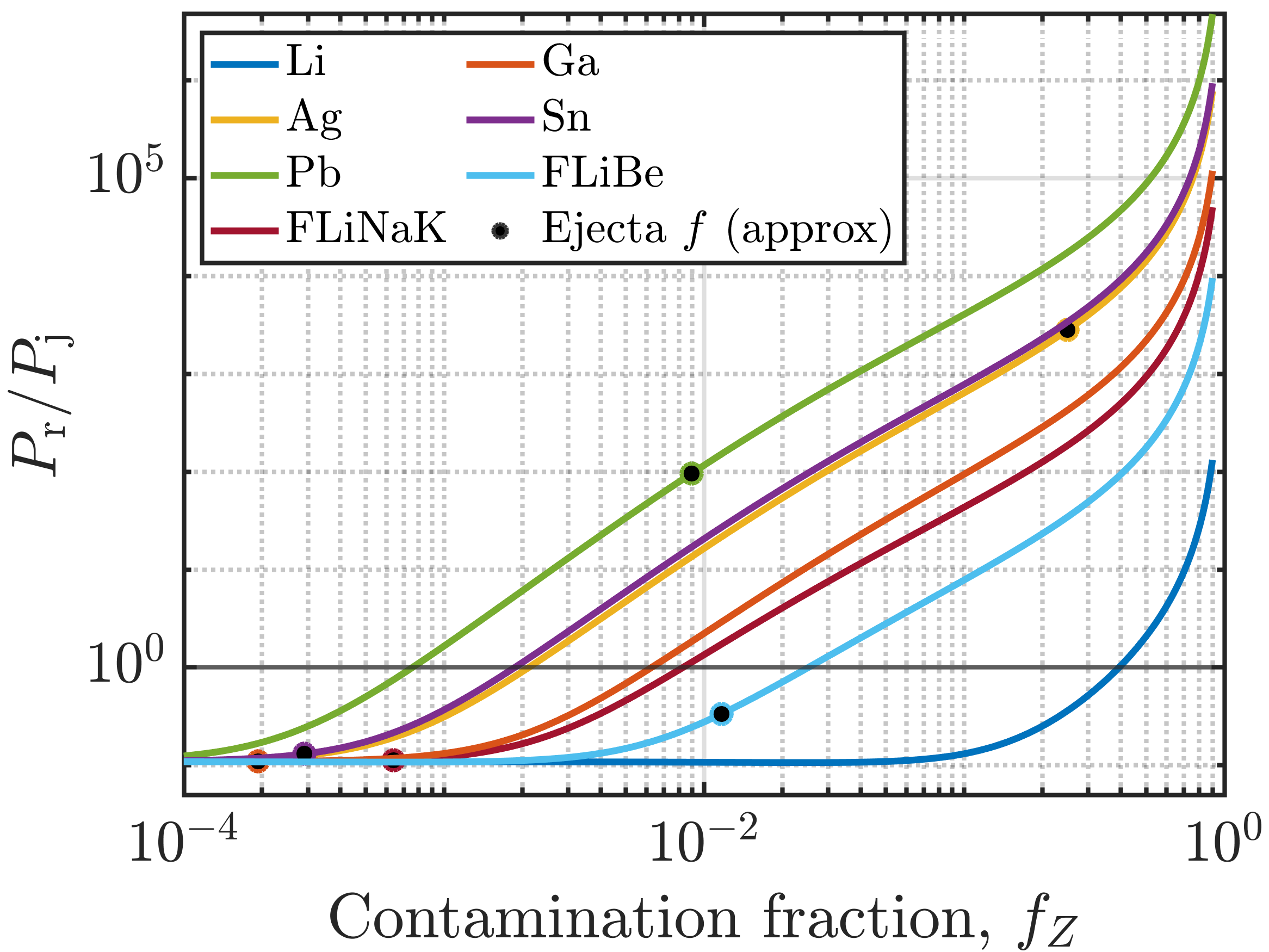}
    \caption{Normalized radiated power ($P_{\mathrm{r}}$) by the Ohmic heating power ($P_{\mathrm{j}}$) from/to a z-pinch fusion core plasma after introduction of contamination. As the normalized power exceeds unity, the radiated power cools the core and reduces the likelihood of fusion events. The impurity selection is consistent with materials frequently researched for compatibility as a LM PFC. A marker is placed at the projected contamination fraction for the candidate ejecta, listed in table \ref{tab:cont_prob_cand}.}
    \label{fig:allow_cont}
\end{figure}

To understand the effect of increased contamination in a pinch, the normalized power loss ($P_{\mathrm{r}}/P_{\mathrm{j}}$) is estimated through a range of contamination levels, expressed as a ratio of the contaminant ion density to the total density,

\begin{equation}\label{eq:cont_fract}
    f_Z = \frac{n_{Z}}{n_{\mathrm{tot}}}.
\end{equation}

Both variables in the normalized power loss are affected by $f_Z$ differently and, as a result, each candidate reaches unity at different contamination fractions, illustrated in Figure \ref{fig:allow_cont}.

When no contamination or very small amounts of contamination are added to the plasma, ohmic heating dominates radiation by one order of magnitude. With increasing $f_Z$, the radiation losses, however, quickly dominate and determine the limits for allowable contamination, taken to be the maximum fraction for power balance. These results are dramatically different from limits previously calculated for tokamaks \cite{critical_exploration}. In this study, a limit of the effective mean charge state ($\langle Z_{\mathrm{eff}}\rangle \leq 2.5$) was reached from the scaling of the lower hybrid, ion cyclotron, and electron cyclotron current drive efficiencies because those are the main sources of power in to a tokamak. The higher allowable impurity charge state correlates to a higher allowable contamination fraction for those devices. This illustrates a possible resiliency to higher contamination levels which exists in steady-state configurations.

\begin{table*}[ht]
    \centering
    \begin{tabular}{|c|c|c|c|c|c|c|}
        \hline
        Candidate, & Estimated Impurity & Contamination & Effective Charge & Allowable Impurity & Radiation Time \\
        $Z$ & Charge State, $\langle Z_{\mathrm{con}} \rangle$ & Fraction, $f_Z$ & State $\langle Z_{\mathrm{eff}}\rangle$ & Mass [\unit{\micro\gram}] & Constant [\unit{\micro\second}]\\
        \hline
        Li & \num{2.99} & \num{0.3936} & \num{1.79} & \num{84.59} & \num{25.06} \\
        Ga & \num{28.42}$^\dagger$ & \num{0.0063} & \num{1.17} & \num{8.27} & \num{17.27} \\
        Ag & \num{36.09} & \num{0.0021} & \num{1.07} & \num{4.30} & \num{17.82} \\
        Sn & \num{38.36} & \num{0.0019} & \num{1.07} & \num{4.18} & \num{17.85} \\
        Pb & \num{45.26}$^\dagger$ & \num{0.0008} & \num{1.03} & \num{2.95} & \num{18.10} \\
        FLiBe & 33.99 & 0.0252 & 1.83 & 35.36 & 15.07 \\
        FLiNaK & 47.48 & 0.0083 & 1.39 & 19.79 & 16.23 \\
        \hline
    \end{tabular}\\
    $^\dagger$Data extrapolated from nearest atomic numbers listed in \cite{coronal_equilibrium}.
    \vspace{-2mm}
    \caption{Allowable contamination for candidate LM PFC materials based on the Pease-Braginskii equilibrium ($P_{\mathrm{r}}/P_{\mathrm{j}} = 1$). Information listed is based on completely hydrogenic core plasmas and a single element contamination or equal molar ratio contamination from homogeneous mixtures. Core temperature, number density, and radius is based on adiabatic scaling of a Q=1, z-pinch plasma.}
    \label{tab:contamination}
\end{table*}

 Using the analysis developed by \citet{radiation_contraction}, a radiation time constant ($\tau_{\mathrm{r}}$) is calculated for the core to collapse,

 \begin{equation}
     \tau_{\mathrm{r}} = 3Nk_{\mathrm{B}}T/P_{\mathrm{r}}.
 \end{equation}

 This time constant describes the time it would take the radiation loss to release all the internal kinetic energy of the plasma, assuming no Ohmic heating. For the contaminated plasmas, the mixed species radiation rates previously estimated are used for this analysis, as well. The time constant calculated and shown in Table \ref{tab:contamination} assumes the plasma began at breakeven conditions. Since z-pinches are expected with lifetimes on the order of tens of microseconds, the radiative collapse of a contaminated core beyond $P_{\mathrm{r}}/P_{\mathrm{j}}=1$ may not completely dissipate the core energy. Although, as the radiated energy scales very rapidly with contamination fraction, it would not take much more and,in some cases, one droplet of ejecta has enough mass to elevate the contamination fraction well beyond the $P_{\mathrm{r}}/P_{\mathrm{j}}$ unity.

 \section{Conclusions}

Based on experimental findings, a proportionality is noted between the annular jet's vertical velocity and the applied magnetic pressure. Also, small diameter ejecta are found to travel at higher velocities, between \num{-10} to \num{15}~\unit{\meter/\second}, with both radial and vertical components to its trajectory. In the scope of z-pinch reactor designs, neither pose a threat to the pinch which produced the response in the liquid, but do to a subsequent pinch in a repetitively pulsed device. Based on the jet movement in LEX, a time frame between \num{100}--\num{120}~\unit{\milli\second} may be suitable for pulse repetition more rapid than fluid settling time.

A probability of 4.6\% is calculated for the likelihood of a droplet forming within the velocity space to end up within the volume of a subsequent pinch. A mass of \num{48.85}~\unit{\micro\gram} is calculated as the most likely to form for a droplet. These both suggest a relatively high likelihood of potentially severe contamination.

Using the ratio of the viscosity, density, and surface tension between LM PFC candidates and our tin-bismuth surrogate, estimates are made about the potential mass and velocity distributions for these candidates' ejecta. Finally, contamination thresholds are calculated for each of the candidates using a normalized power balance for a z-pinch. The estimated droplet masses are compared to the threshold and a time for radiative collapse of the pinch is calculated at the threshold.

These findings demonstrate an elimination from using large amounts of certain high-Z material for use as LM PFCs in z-pinch reactors, to include silver and lead. From the strong LM behavior, there is a high likelihood of core contamination and the amount of energy radiated away from the core after introduction of such impurities would leech large amounts energy quickly, preventing the fusion of hydrogenic ions. Other candidate materials, such as lithium, FLiBe, FLiNaK, gallium, and tin may not cause contamination as severe and should still be considered for use.

\begin{acknowledgments}
    This research is supported by the Department of Energy's Advanced Research Projects Agency--Energy through the BETHE program, under grant DE--AR0001263.
\end{acknowledgments}

\bibliography{references}

\providecommand{\noopsort}[1]{}\providecommand{\singleletter}[1]{#1}%
\begin{thebibliography}{41}%
\makeatletter
\providecommand \@ifxundefined [1]{%
 \@ifx{#1\undefined}
}%
\providecommand \@ifnum [1]{%
 \ifnum #1\expandafter \@firstoftwo
 \else \expandafter \@secondoftwo
 \fi
}%
\providecommand \@ifx [1]{%
 \ifx #1\expandafter \@firstoftwo
 \else \expandafter \@secondoftwo
 \fi
}%
\providecommand \natexlab [1]{#1}%
\providecommand \enquote  [1]{``#1''}%
\providecommand \bibnamefont  [1]{#1}%
\providecommand \bibfnamefont [1]{#1}%
\providecommand \citenamefont [1]{#1}%
\providecommand \href@noop [0]{\@secondoftwo}%
\providecommand \href [0]{\begingroup \@sanitize@url \@href}%
\providecommand \@href[1]{\@@startlink{#1}\@@href}%
\providecommand \@@href[1]{\endgroup#1\@@endlink}%
\providecommand \@sanitize@url [0]{\catcode `\\12\catcode `\$12\catcode
  `\&12\catcode `\#12\catcode `\^12\catcode `\_12\catcode `\%12\relax}%
\providecommand \@@startlink[1]{}%
\providecommand \@@endlink[0]{}%
\providecommand \url  [0]{\begingroup\@sanitize@url \@url }%
\providecommand \@url [1]{\endgroup\@href {#1}{\urlprefix }}%
\providecommand \urlprefix  [0]{URL }%
\providecommand \Eprint [0]{\href }%
\providecommand \doibase [0]{https://doi.org/}%
\providecommand \selectlanguage [0]{\@gobble}%
\providecommand \bibinfo  [0]{\@secondoftwo}%
\providecommand \bibfield  [0]{\@secondoftwo}%
\providecommand \translation [1]{[#1]}%
\providecommand \BibitemOpen [0]{}%
\providecommand \bibitemStop [0]{}%
\providecommand \bibitemNoStop [0]{.\EOS\space}%
\providecommand \EOS [0]{\spacefactor3000\relax}%
\providecommand \BibitemShut  [1]{\csname bibitem#1\endcsname}%
\let\auto@bib@innerbib\@empty
\bibitem [{\citenamefont {del Rio}\ \emph {et~al.}(2020)\citenamefont {del
  Rio}, \citenamefont {Gautam},\ and\ \citenamefont
  {Carter}}]{deuterium_addition_LiSn}%
  \BibitemOpen
  \bibfield  {author} {\bibinfo {author} {\bibfnamefont {B.~G.}\ \bibnamefont
  {del Rio}}, \bibinfo {author} {\bibfnamefont {G.~S.}\ \bibnamefont
  {Gautam}},\ and\ \bibinfo {author} {\bibfnamefont {E.~A.}\ \bibnamefont
  {Carter}},\ }\bibfield  {title} {\bibinfo {title} {Deuterium addition to
  liquid {Li-Sn} alloys: implications for plasma-facing applications},\
  }\href@noop {} {\bibfield  {journal} {\bibinfo  {journal} {Nuclear Fusion}\
  }\textbf {\bibinfo {volume} {60}} (\bibinfo {year} {2020})}\BibitemShut
  {NoStop}%
\bibitem [{\citenamefont {Rubel}(2019)}]{trit_breed_impact}%
  \BibitemOpen
  \bibfield  {author} {\bibinfo {author} {\bibfnamefont {M.}~\bibnamefont
  {Rubel}},\ }\bibfield  {title} {\bibinfo {title} {Fusion neutrons: {T}ritium
  breeding and impact on wall materials and components of diagnostic systems},\
  }\href@noop {} {\bibfield  {journal} {\bibinfo  {journal} {Journal of Fusion
  Energy}\ }\textbf {\bibinfo {volume} {38}},\ \bibinfo {pages} {315} (\bibinfo
  {year} {2019})}\BibitemShut {NoStop}%
\bibitem [{\citenamefont {Cao}\ \emph {et~al.}(2014)\citenamefont {Cao},
  \citenamefont {Ou}, \citenamefont {Tian}, \citenamefont {Wang}, \citenamefont
  {Zhu}, \citenamefont {Wang}, \citenamefont {Gou}, \citenamefont {Yang},\ and\
  \citenamefont {Chen}}]{new_facility_Flili}%
  \BibitemOpen
  \bibfield  {author} {\bibinfo {author} {\bibfnamefont {X.}~\bibnamefont
  {Cao}}, \bibinfo {author} {\bibfnamefont {W.}~\bibnamefont {Ou}}, \bibinfo
  {author} {\bibfnamefont {S.}~\bibnamefont {Tian}}, \bibinfo {author}
  {\bibfnamefont {C.}~\bibnamefont {Wang}}, \bibinfo {author} {\bibfnamefont
  {Z.}~\bibnamefont {Zhu}}, \bibinfo {author} {\bibfnamefont {J.}~\bibnamefont
  {Wang}}, \bibinfo {author} {\bibfnamefont {F.}~\bibnamefont {Gou}}, \bibinfo
  {author} {\bibfnamefont {D.}~\bibnamefont {Yang}},\ and\ \bibinfo {author}
  {\bibfnamefont {S.}~\bibnamefont {Chen}},\ }\bibfield  {title} {\bibinfo
  {title} {A new facility for studying plasma interacting with flowing liquid
  lithium surface},\ }\href@noop {} {\bibfield  {journal} {\bibinfo  {journal}
  {Fusion Engineering and Design}\ }\textbf {\bibinfo {volume} {89}},\ \bibinfo
  {pages} {2864} (\bibinfo {year} {2014})}\BibitemShut {NoStop}%
\bibitem [{\citenamefont {Kaita}\ \emph {et~al.}(2017)\citenamefont {Kaita},
  \citenamefont {Lucia}, \citenamefont {Allain}, \citenamefont {Bedoya},
  \citenamefont {Bell}, \citenamefont {Boyle}, \citenamefont {Capece},
  \citenamefont {Jaworski}, \citenamefont {Koel}, \citenamefont {Majeski},
  \citenamefont {Roszell}, \citenamefont {Schmitt}, \citenamefont {Scotti},
  \citenamefont {Skinner},\ and\ \citenamefont
  {Soukhanovskii}}]{hyd_retent_implications_high_z}%
  \BibitemOpen
  \bibfield  {author} {\bibinfo {author} {\bibfnamefont {R.}~\bibnamefont
  {Kaita}}, \bibinfo {author} {\bibfnamefont {M.}~\bibnamefont {Lucia}},
  \bibinfo {author} {\bibfnamefont {J.~P.}\ \bibnamefont {Allain}}, \bibinfo
  {author} {\bibfnamefont {D.}~\bibnamefont {Bedoya}}, \bibinfo {author}
  {\bibfnamefont {R.}~\bibnamefont {Bell}}, \bibinfo {author} {\bibfnamefont
  {D.}~\bibnamefont {Boyle}}, \bibinfo {author} {\bibfnamefont
  {A.}~\bibnamefont {Capece}}, \bibinfo {author} {\bibfnamefont
  {M.}~\bibnamefont {Jaworski}}, \bibinfo {author} {\bibfnamefont {B.~E.}\
  \bibnamefont {Koel}}, \bibinfo {author} {\bibfnamefont {R.}~\bibnamefont
  {Majeski}}, \bibinfo {author} {\bibfnamefont {J.}~\bibnamefont {Roszell}},
  \bibinfo {author} {\bibfnamefont {J.}~\bibnamefont {Schmitt}}, \bibinfo
  {author} {\bibfnamefont {F.}~\bibnamefont {Scotti}}, \bibinfo {author}
  {\bibfnamefont {C.~H.}\ \bibnamefont {Skinner}},\ and\ \bibinfo {author}
  {\bibfnamefont {V.}~\bibnamefont {Soukhanovskii}},\ }\bibfield  {title}
  {\bibinfo {title} {Hydrogen retention in lithium on metallic walls from
  \emph{in vacuo} analysis in {LTX} and implications for high-{Z} plasma-facing
  components in {NSTX-U}},\ }\href@noop {} {\bibfield  {journal} {\bibinfo
  {journal} {Fusion Engineering and Design}\ }\textbf {\bibinfo {volume}
  {117}},\ \bibinfo {pages} {135} (\bibinfo {year} {2017})}\BibitemShut
  {NoStop}%
\bibitem [{\citenamefont {Kondo}\ \emph {et~al.}(2012)\citenamefont {Kondo},
  \citenamefont {Oshima}, \citenamefont {Tanaka}, \citenamefont {Muroga},\ and\
  \citenamefont {Sagara}}]{Gas_recycling_bubbling}%
  \BibitemOpen
  \bibfield  {author} {\bibinfo {author} {\bibfnamefont {M.}~\bibnamefont
  {Kondo}}, \bibinfo {author} {\bibfnamefont {T.}~\bibnamefont {Oshima}},
  \bibinfo {author} {\bibfnamefont {M.}~\bibnamefont {Tanaka}}, \bibinfo
  {author} {\bibfnamefont {T.}~\bibnamefont {Muroga}},\ and\ \bibinfo {author}
  {\bibfnamefont {A.}~\bibnamefont {Sagara}},\ }\bibfield  {title} {\bibinfo
  {title} {Hydrogen transport through interface between gas bubbling and liquid
  breeders},\ }\href@noop {} {\bibfield  {journal} {\bibinfo  {journal} {Fusion
  Engineering and Design}\ }\textbf {\bibinfo {volume} {87}},\ \bibinfo {pages}
  {1788} (\bibinfo {year} {2012})}\BibitemShut {NoStop}%
\bibitem [{\citenamefont {Marenkov}\ and\ \citenamefont
  {Pshenov}(2020)}]{vapor_shielding}%
  \BibitemOpen
  \bibfield  {author} {\bibinfo {author} {\bibfnamefont {E.}~\bibnamefont
  {Marenkov}}\ and\ \bibinfo {author} {\bibfnamefont {A.}~\bibnamefont
  {Pshenov}},\ }\bibfield  {title} {\bibinfo {title} {Vapor shielding of liquid
  lithium divertor target during steady state and transient events},\
  }\href@noop {} {\bibfield  {journal} {\bibinfo  {journal} {Nuclear Fusion}\
  }\textbf {\bibinfo {volume} {60}} (\bibinfo {year} {2020})}\BibitemShut
  {NoStop}%
\bibitem [{\citenamefont {Rindt}\ \emph {et~al.}(2019)\citenamefont {Rindt},
  \citenamefont {Morgan}, \citenamefont {van Eden},\ and\ \citenamefont
  {Cardozo}}]{Li_divertor_design}%
  \BibitemOpen
  \bibfield  {author} {\bibinfo {author} {\bibfnamefont {P.}~\bibnamefont
  {Rindt}}, \bibinfo {author} {\bibfnamefont {T.~W.}\ \bibnamefont {Morgan}},
  \bibinfo {author} {\bibfnamefont {G.~G.}\ \bibnamefont {van Eden}},\ and\
  \bibinfo {author} {\bibfnamefont {N.~J.~L.}\ \bibnamefont {Cardozo}},\
  }\bibfield  {title} {\bibinfo {title} {Power handling and vapor shielding of
  pre-filled lithium divertor targets in {Magnum-PSI}},\ }\href@noop {}
  {\bibfield  {journal} {\bibinfo  {journal} {Nuclear Fusion}\ }\textbf
  {\bibinfo {volume} {59}} (\bibinfo {year} {2019})}\BibitemShut {NoStop}%
\bibitem [{\citenamefont {van Eden}\ \emph {et~al.}(2016)\citenamefont {van
  Eden}, \citenamefont {Morgan}, \citenamefont {Aussems}, \citenamefont
  {van~den Berg}, \citenamefont {Bystrov},\ and\ \citenamefont {van~de
  Sanden}}]{Sn_Vapor_Shielding}%
  \BibitemOpen
  \bibfield  {author} {\bibinfo {author} {\bibfnamefont {G.~G.}\ \bibnamefont
  {van Eden}}, \bibinfo {author} {\bibfnamefont {T.}~\bibnamefont {Morgan}},
  \bibinfo {author} {\bibfnamefont {D.~U.~B.}\ \bibnamefont {Aussems}},
  \bibinfo {author} {\bibfnamefont {M.~A.}\ \bibnamefont {van~den Berg}},
  \bibinfo {author} {\bibfnamefont {K.}~\bibnamefont {Bystrov}},\ and\ \bibinfo
  {author} {\bibfnamefont {M.~C.~M.}\ \bibnamefont {van~de Sanden}},\
  }\bibfield  {title} {\bibinfo {title} {Self-regulated plasma heat flux
  mitigation due to liquid {Sn} vapor shielding},\ }\href@noop {} {\bibfield
  {journal} {\bibinfo  {journal} {Physical Review Letters}\ }\textbf {\bibinfo
  {volume} {116}} (\bibinfo {year} {2016})}\BibitemShut {NoStop}%
\bibitem [{\citenamefont {Ji}\ \emph {et~al.}(2005)\citenamefont {Ji},
  \citenamefont {Fox}, \citenamefont {Pace},\ and\ \citenamefont
  {Rappaport}}]{small_MHD_waves}%
  \BibitemOpen
  \bibfield  {author} {\bibinfo {author} {\bibfnamefont {H.}~\bibnamefont
  {Ji}}, \bibinfo {author} {\bibfnamefont {W.}~\bibnamefont {Fox}}, \bibinfo
  {author} {\bibfnamefont {D.}~\bibnamefont {Pace}},\ and\ \bibinfo {author}
  {\bibfnamefont {H.~L.}\ \bibnamefont {Rappaport}},\ }\bibfield  {title}
  {\bibinfo {title} {Study of small-amplitude magnetohydrodynamic surface waves
  on liquid metal},\ }\href@noop {} {\bibfield  {journal} {\bibinfo  {journal}
  {Physics of Plasmas}\ }\textbf {\bibinfo {volume} {12}} (\bibinfo {year}
  {2005})}\BibitemShut {NoStop}%
\bibitem [{\citenamefont {Buryak}\ \emph {et~al.}(2012)\citenamefont {Buryak},
  \citenamefont {Kolesnichenko}, \citenamefont {Kolesmchenko},\ and\
  \citenamefont {Smolentsev}}]{free-surface_MHD_Heat}%
  \BibitemOpen
  \bibfield  {author} {\bibinfo {author} {\bibfnamefont {V.~V.}\ \bibnamefont
  {Buryak}}, \bibinfo {author} {\bibfnamefont {A.}~\bibnamefont
  {Kolesnichenko}}, \bibinfo {author} {\bibfnamefont {A.}~\bibnamefont
  {Kolesmchenko}},\ and\ \bibinfo {author} {\bibfnamefont {S.}~\bibnamefont
  {Smolentsev}},\ }\bibfield  {title} {\bibinfo {title} {Free-surface {MHD}
  flows as a potential tool for high heat flux removal in fusion
  applications},\ }\href@noop {} {\bibfield  {journal} {\bibinfo  {journal}
  {Magnetohydrodynamics}\ }\textbf {\bibinfo {volume} {48}},\ \bibinfo {pages}
  {651} (\bibinfo {year} {2012})}\BibitemShut {NoStop}%
\bibitem [{\citenamefont {Jaworski}\ \emph {et~al.}(2013)\citenamefont
  {Jaworski}, \citenamefont {Khodak},\ and\ \citenamefont
  {Kaita}}]{nstx_lld_li_pfc}%
  \BibitemOpen
  \bibfield  {author} {\bibinfo {author} {\bibfnamefont {M.}~\bibnamefont
  {Jaworski}}, \bibinfo {author} {\bibfnamefont {A.}~\bibnamefont {Khodak}},\
  and\ \bibinfo {author} {\bibfnamefont {R.}~\bibnamefont {Kaita}},\ }\bibfield
   {title} {\bibinfo {title} {Liquid-metal plasma-facing component research on
  the {N}ational {S}pherical {T}orus {E}xperiment},\ }\href@noop {} {\bibfield
  {journal} {\bibinfo  {journal} {Plasma Physics and Controlled Fusion}\
  }\textbf {\bibinfo {volume} {55}} (\bibinfo {year} {2013})}\BibitemShut
  {NoStop}%
\bibitem [{\citenamefont {Andruczyk}\ \emph {et~al.}(2022)\citenamefont
  {Andruczyk}, \citenamefont {Shone}, \citenamefont {Koyn}, ,\ and\
  \citenamefont {Allain}}]{Li_HIDRA}%
  \BibitemOpen
  \bibfield  {author} {\bibinfo {author} {\bibfnamefont {D.}~\bibnamefont
  {Andruczyk}}, \bibinfo {author} {\bibfnamefont {A.}~\bibnamefont {Shone}},
  \bibinfo {author} {\bibfnamefont {Z.}~\bibnamefont {Koyn}}, ,\ and\ \bibinfo
  {author} {\bibfnamefont {J.~P.}\ \bibnamefont {Allain}},\ }\bibfield  {title}
  {\bibinfo {title} {First lithium experiments in {HIDRA} and evidence of
  helium retention during quasi-steady-state stellarator plasma operations},\
  }\href@noop {} {\bibfield  {journal} {\bibinfo  {journal} {Plasma Physics and
  Controlled Fusion}\ }\textbf {\bibinfo {volume} {64}} (\bibinfo {year}
  {2022})}\BibitemShut {NoStop}%
\bibitem [{\citenamefont {Tabar\'es}\ \emph {et~al.}(2017)\citenamefont
  {Tabar\'es}, \citenamefont {Oyarzabal}, \citenamefont {Martin-Rojo},
  \citenamefont {Tafalla}, \citenamefont {de~Castro}, \citenamefont {Medina},
  \citenamefont {Ochando}, \citenamefont {Zurro}, \citenamefont {McCarthy},\
  and\ \citenamefont {the TJ-II~Team}}]{LiSn_divertor}%
  \BibitemOpen
  \bibfield  {author} {\bibinfo {author} {\bibfnamefont {F.}~\bibnamefont
  {Tabar\'es}}, \bibinfo {author} {\bibfnamefont {E.}~\bibnamefont
  {Oyarzabal}}, \bibinfo {author} {\bibfnamefont {A.}~\bibnamefont
  {Martin-Rojo}}, \bibinfo {author} {\bibfnamefont {D.}~\bibnamefont
  {Tafalla}}, \bibinfo {author} {\bibfnamefont {A.}~\bibnamefont {de~Castro}},
  \bibinfo {author} {\bibfnamefont {F.}~\bibnamefont {Medina}}, \bibinfo
  {author} {\bibfnamefont {M.}~\bibnamefont {Ochando}}, \bibinfo {author}
  {\bibfnamefont {B.}~\bibnamefont {Zurro}}, \bibinfo {author} {\bibfnamefont
  {K.}~\bibnamefont {McCarthy}},\ and\ \bibinfo {author} {\bibnamefont {the
  TJ-II~Team}},\ }\bibfield  {title} {\bibinfo {title} {Experimental tests of
  {LiSn} alloys as potential liquid metal for the divertor target in a fusion
  reactor},\ }\href@noop {} {\bibfield  {journal} {\bibinfo  {journal} {Nuclear
  Materials and Energy}\ }\textbf {\bibinfo {volume} {12}},\ \bibinfo {pages}
  {1368} (\bibinfo {year} {2017})}\BibitemShut {NoStop}%
\bibitem [{\citenamefont {Loureiro}\ \emph {et~al.}(2017)\citenamefont
  {Loureiro}, \citenamefont {Tabar\'es}, \citenamefont {Fernandes},
  \citenamefont {Silva}, \citenamefont {Gomes}, \citenamefont {Alves},
  \citenamefont {Mateus}, \citenamefont {Pereira}, \citenamefont {Alves},\ and\
  \citenamefont {Figueiredo}}]{lisn_pfc_behavior}%
  \BibitemOpen
  \bibfield  {author} {\bibinfo {author} {\bibfnamefont {J.}~\bibnamefont
  {Loureiro}}, \bibinfo {author} {\bibfnamefont {F.}~\bibnamefont {Tabar\'es}},
  \bibinfo {author} {\bibfnamefont {H.}~\bibnamefont {Fernandes}}, \bibinfo
  {author} {\bibfnamefont {C.}~\bibnamefont {Silva}}, \bibinfo {author}
  {\bibfnamefont {R.}~\bibnamefont {Gomes}}, \bibinfo {author} {\bibfnamefont
  {E.}~\bibnamefont {Alves}}, \bibinfo {author} {\bibfnamefont
  {R.}~\bibnamefont {Mateus}}, \bibinfo {author} {\bibfnamefont
  {T.}~\bibnamefont {Pereira}}, \bibinfo {author} {\bibfnamefont
  {H.}~\bibnamefont {Alves}},\ and\ \bibinfo {author} {\bibfnamefont
  {H.}~\bibnamefont {Figueiredo}},\ }\bibfield  {title} {\bibinfo {title}
  {Behavior of liquid {Li-Sn} alloy as plasma facing material on {ISTTOK}},\
  }\href@noop {} {\bibfield  {journal} {\bibinfo  {journal} {Fusion Engineering
  and Design}\ }\textbf {\bibinfo {volume} {117}},\ \bibinfo {pages} {208}
  (\bibinfo {year} {2017})}\BibitemShut {NoStop}%
\bibitem [{\citenamefont {Kessel}\ \emph {et~al.}(2019)\citenamefont {Kessel},
  \citenamefont {Andruczyk}, \citenamefont {Blanchard}, \citenamefont {Bohm},
  \citenamefont {Davis}, \citenamefont {Hollis}, \citenamefont {Humrickhouse},
  \citenamefont {Hvasta}, \citenamefont {Jaworski}, \citenamefont {Jun},
  \citenamefont {Katoh}, \citenamefont {Khodak}, \citenamefont {Klein},
  \citenamefont {Kolemen}, \citenamefont {Larsen}, \citenamefont {Majeski},
  \citenamefont {Merrill}, \citenamefont {Morley}, \citenamefont {Neilson},
  \citenamefont {Pint}, \citenamefont {Rensink}, \citenamefont {Rognlien},
  \citenamefont {Rowcliffe}, \citenamefont {Smolentsev}, \citenamefont
  {Tillack}, \citenamefont {Waganer}, \citenamefont {Wallace}, \citenamefont
  {Wilson},\ and\ \citenamefont {Yoon}}]{critical_exploration}%
  \BibitemOpen
  \bibfield  {author} {\bibinfo {author} {\bibfnamefont {C.~E.}\ \bibnamefont
  {Kessel}}, \bibinfo {author} {\bibfnamefont {D.}~\bibnamefont {Andruczyk}},
  \bibinfo {author} {\bibfnamefont {J.~P.}\ \bibnamefont {Blanchard}}, \bibinfo
  {author} {\bibfnamefont {T.}~\bibnamefont {Bohm}}, \bibinfo {author}
  {\bibfnamefont {A.}~\bibnamefont {Davis}}, \bibinfo {author} {\bibfnamefont
  {K.}~\bibnamefont {Hollis}}, \bibinfo {author} {\bibfnamefont {P.~W.}\
  \bibnamefont {Humrickhouse}}, \bibinfo {author} {\bibfnamefont
  {M.}~\bibnamefont {Hvasta}}, \bibinfo {author} {\bibfnamefont
  {M.}~\bibnamefont {Jaworski}}, \bibinfo {author} {\bibfnamefont
  {J.}~\bibnamefont {Jun}}, \bibinfo {author} {\bibfnamefont {Y.}~\bibnamefont
  {Katoh}}, \bibinfo {author} {\bibfnamefont {A.}~\bibnamefont {Khodak}},
  \bibinfo {author} {\bibfnamefont {J.}~\bibnamefont {Klein}}, \bibinfo
  {author} {\bibfnamefont {E.}~\bibnamefont {Kolemen}}, \bibinfo {author}
  {\bibfnamefont {G.}~\bibnamefont {Larsen}}, \bibinfo {author} {\bibfnamefont
  {R.}~\bibnamefont {Majeski}}, \bibinfo {author} {\bibfnamefont {B.~J.}\
  \bibnamefont {Merrill}}, \bibinfo {author} {\bibfnamefont {N.~B.}\
  \bibnamefont {Morley}}, \bibinfo {author} {\bibfnamefont {G.~H.}\
  \bibnamefont {Neilson}}, \bibinfo {author} {\bibfnamefont {B.}~\bibnamefont
  {Pint}}, \bibinfo {author} {\bibfnamefont {M.~E.}\ \bibnamefont {Rensink}},
  \bibinfo {author} {\bibfnamefont {T.~D.}\ \bibnamefont {Rognlien}}, \bibinfo
  {author} {\bibfnamefont {A.~F.}\ \bibnamefont {Rowcliffe}}, \bibinfo {author}
  {\bibfnamefont {S.}~\bibnamefont {Smolentsev}}, \bibinfo {author}
  {\bibfnamefont {M.~S.}\ \bibnamefont {Tillack}}, \bibinfo {author}
  {\bibfnamefont {L.~M.}\ \bibnamefont {Waganer}}, \bibinfo {author}
  {\bibfnamefont {G.~M.}\ \bibnamefont {Wallace}}, \bibinfo {author}
  {\bibfnamefont {P.}~\bibnamefont {Wilson}},\ and\ \bibinfo {author}
  {\bibfnamefont {S.-J.}\ \bibnamefont {Yoon}},\ }\bibfield  {title} {\bibinfo
  {title} {Critical exploration of liquid metal plasma-facing components in a
  {F}usion {N}uclear {S}cience {F}acility},\ }\href@noop {} {\bibfield
  {journal} {\bibinfo  {journal} {Fusion Science and Technology}\ }\textbf
  {\bibinfo {volume} {75}},\ \bibinfo {pages} {886} (\bibinfo {year}
  {2019})}\BibitemShut {NoStop}%
\bibitem [{\citenamefont {Gohar}\ and\ \citenamefont
  {Smith}(2000)}]{neutron_multipliers}%
  \BibitemOpen
  \bibfield  {author} {\bibinfo {author} {\bibfnamefont {Y.}~\bibnamefont
  {Gohar}}\ and\ \bibinfo {author} {\bibfnamefont {D.}~\bibnamefont {Smith}},\
  }\href@noop {} {\emph {\bibinfo {title} {Multiplier, moderator, and reflector
  materials for advanced lithium-vanadium fusion blankets}}},\ \bibinfo {type}
  {Tech. Rep.}\ \bibinfo {number} {ANL/TD/CP-98461}\ (\bibinfo  {institution}
  {Argonne National Laboratory},\ \bibinfo {year} {2000})\BibitemShut {NoStop}%
\bibitem [{\citenamefont {Morgan}\ \emph {et~al.}(2015)\citenamefont {Morgan},
  \citenamefont {van~den Bekerom},\ and\ \citenamefont
  {Temmerman}}]{sn_cps_iter}%
  \BibitemOpen
  \bibfield  {author} {\bibinfo {author} {\bibfnamefont {T.}~\bibnamefont
  {Morgan}}, \bibinfo {author} {\bibfnamefont {D.}~\bibnamefont {van~den
  Bekerom}},\ and\ \bibinfo {author} {\bibfnamefont {G.~D.}\ \bibnamefont
  {Temmerman}},\ }\bibfield  {title} {\bibinfo {title} {Interaction of a
  tin-based capillary porous structure with {ITER/DEMO} relevant plasma
  conditions},\ }\href@noop {} {\bibfield  {journal} {\bibinfo  {journal}
  {Journal of Nuclear Materials}\ }\textbf {\bibinfo {volume} {463}},\ \bibinfo
  {pages} {1256} (\bibinfo {year} {2015})}\BibitemShut {NoStop}%
\bibitem [{\citenamefont {Mirnov}\ and\ \citenamefont
  {Evtikhin}(2006)}]{Ga_Li_Divertor}%
  \BibitemOpen
  \bibfield  {author} {\bibinfo {author} {\bibfnamefont {S.}~\bibnamefont
  {Mirnov}}\ and\ \bibinfo {author} {\bibfnamefont {V.}~\bibnamefont
  {Evtikhin}},\ }\bibfield  {title} {\bibinfo {title} {The tests of liquid
  metals ({Ga}, {Li}) as plasma facing components in {T-3M} and {T-11M}
  tokamaks},\ }\href@noop {} {\bibfield  {journal} {\bibinfo  {journal} {Fusion
  Engineering and Design}\ }\textbf {\bibinfo {volume} {81}},\ \bibinfo {pages}
  {113} (\bibinfo {year} {2006})}\BibitemShut {NoStop}%
\bibitem [{\citenamefont {Hutchinson}(2002)}]{hutch}%
  \BibitemOpen
  \bibfield  {author} {\bibinfo {author} {\bibfnamefont {I.~H.}\ \bibnamefont
  {Hutchinson}},\ }\href@noop {} {\emph {\bibinfo {title} {Principles of Plasma
  Diagnostics}}},\ \bibinfo {edition} {2nd}\ ed.\ (\bibinfo  {publisher}
  {Cambridge University Press},\ \bibinfo {year} {2002})\BibitemShut {NoStop}%
\bibitem [{\citenamefont {Faltinsen}(2005)}]{odd}%
  \BibitemOpen
  \bibfield  {author} {\bibinfo {author} {\bibfnamefont {O.~M.}\ \bibnamefont
  {Faltinsen}},\ }\href@noop {} {\emph {\bibinfo {title} {Hydrodynamics of
  High-Speed Marine Vehicles}}},\ \bibinfo {edition} {1st}\ ed.\ (\bibinfo
  {publisher} {Cambridge University Press},\ \bibinfo {year}
  {2005})\BibitemShut {NoStop}%
\bibitem [{\citenamefont {Levitt}\ \emph {et~al.}(2023)\citenamefont {Levitt},
  \citenamefont {Meier}, \citenamefont {Umstattd}, \citenamefont {Barhydt},
  \citenamefont {Datta}, \citenamefont {Liekhus-Schmaltz}, \citenamefont
  {Sutherland},\ and\ \citenamefont {Nelson}}]{zap_approach}%
  \BibitemOpen
  \bibfield  {author} {\bibinfo {author} {\bibfnamefont {B.}~\bibnamefont
  {Levitt}}, \bibinfo {author} {\bibfnamefont {E.~T.}\ \bibnamefont {Meier}},
  \bibinfo {author} {\bibfnamefont {R.}~\bibnamefont {Umstattd}}, \bibinfo
  {author} {\bibfnamefont {J.~R.}\ \bibnamefont {Barhydt}}, \bibinfo {author}
  {\bibfnamefont {I.~A.~M.}\ \bibnamefont {Datta}}, \bibinfo {author}
  {\bibfnamefont {C.}~\bibnamefont {Liekhus-Schmaltz}}, \bibinfo {author}
  {\bibfnamefont {D.~A.}\ \bibnamefont {Sutherland}},\ and\ \bibinfo {author}
  {\bibfnamefont {B.~A.}\ \bibnamefont {Nelson}},\ }\bibfield  {title}
  {\bibinfo {title} {{The {Z}ap {E}nergy approach to commercial fusion}},\
  }\href {https://doi.org/10.1063/5.0163361} {\bibfield  {journal} {\bibinfo
  {journal} {Physics of Plasmas}\ }\textbf {\bibinfo {volume} {30}},\ \bibinfo
  {pages} {090603} (\bibinfo {year} {2023})},\ \Eprint
  {https://arxiv.org/abs/https://pubs.aip.org/aip/pop/article-pdf/doi/10.1063/5.0163361/18128756/090603\_1\_5.0163361.pdf}
  {https://pubs.aip.org/aip/pop/article-pdf/doi/10.1063/5.0163361/18128756/090603\_1\_5.0163361.pdf}
  \BibitemShut {NoStop}%
\bibitem [{\citenamefont {Shumlak}(2020)}]{shumlak_z_pinch}%
  \BibitemOpen
  \bibfield  {author} {\bibinfo {author} {\bibfnamefont {U.}~\bibnamefont
  {Shumlak}},\ }\bibfield  {title} {\bibinfo {title} {Z-pinch fusion},\
  }\href@noop {} {\bibfield  {journal} {\bibinfo  {journal} {Journal of Applied
  Physics}\ }\textbf {\bibinfo {volume} {127}} (\bibinfo {year}
  {2020})}\BibitemShut {NoStop}%
\bibitem [{\citenamefont {Worthington}\ and\ \citenamefont
  {Stewart}(1881)}]{pendant_drop}%
  \BibitemOpen
  \bibfield  {author} {\bibinfo {author} {\bibfnamefont {A.~M.}\ \bibnamefont
  {Worthington}}\ and\ \bibinfo {author} {\bibfnamefont {B.}~\bibnamefont
  {Stewart}},\ }\bibfield  {title} {\bibinfo {title} {Ii. on pendent drops},\
  }\href@noop {} {\bibfield  {journal} {\bibinfo  {journal} {Proceedings of the
  Royal Society of London}\ }\textbf {\bibinfo {volume} {32}},\ \bibinfo
  {pages} {362} (\bibinfo {year} {1881})}\BibitemShut {NoStop}%
\bibitem [{\citenamefont {Dobosz}\ and\ \citenamefont
  {Gancarz}(2018)}]{gamma_mu_snbi}%
  \BibitemOpen
  \bibfield  {author} {\bibinfo {author} {\bibfnamefont {A.}~\bibnamefont
  {Dobosz}}\ and\ \bibinfo {author} {\bibfnamefont {T.}~\bibnamefont
  {Gancarz}},\ }\bibfield  {title} {\bibinfo {title} {Reference data for the
  density, viscosity, and surface tension of liquid {Al?Zn}, {Ag?Sn}, {Bi?Sn},
  {Cu?Sn}, and {Sn?Zn} eutectic alloys},\ }\href@noop {} {\bibfield  {journal}
  {\bibinfo  {journal} {Journal of Physical and Chemical Reference Data}\
  }\textbf {\bibinfo {volume} {47}} (\bibinfo {year} {2018})}\BibitemShut
  {NoStop}%
\bibitem [{\citenamefont {Davison}(1968)}]{mu_li}%
  \BibitemOpen
  \bibfield  {author} {\bibinfo {author} {\bibfnamefont {H.~W.}\ \bibnamefont
  {Davison}},\ }\href@noop {} {\emph {\bibinfo {title} {Compilation of
  Thermophysical Properities of Liquid Lithium}}},\ \bibinfo {type} {Tech.
  Rep.}\ \bibinfo {number} {D-4650}\ (\bibinfo  {institution} {National
  Aeronautics and Space Administration},\ \bibinfo {year} {1968})\BibitemShut
  {NoStop}%
\bibitem [{\citenamefont {Alcock}\ \emph {et~al.}(1984)\citenamefont {Alcock},
  \citenamefont {Itkin},\ and\ \citenamefont {Horrigan}}]{vap_press_metals}%
  \BibitemOpen
  \bibfield  {author} {\bibinfo {author} {\bibfnamefont {C.~B.}\ \bibnamefont
  {Alcock}}, \bibinfo {author} {\bibfnamefont {V.~P.}\ \bibnamefont {Itkin}},\
  and\ \bibinfo {author} {\bibfnamefont {M.~K.}\ \bibnamefont {Horrigan}},\
  }\bibfield  {title} {\bibinfo {title} {Vapour pressure equations for the
  metallic elements: 298–2500k},\ }\href@noop {} {\bibfield  {journal}
  {\bibinfo  {journal} {Canadian Metallurgical Quarterly}\ }\textbf {\bibinfo
  {volume} {23}},\ \bibinfo {pages} {309} (\bibinfo {year} {1984})}\BibitemShut
  {NoStop}%
\bibitem [{\citenamefont {Assael}\ \emph
  {et~al.}(2012{\natexlab{a}})\citenamefont {Assael}, \citenamefont {Armyra},
  \citenamefont {Brillo}, \citenamefont {Stankus}, \citenamefont {Wu},\ and\
  \citenamefont {Wakeham}}]{mu_ga}%
  \BibitemOpen
  \bibfield  {author} {\bibinfo {author} {\bibfnamefont {M.~J.}\ \bibnamefont
  {Assael}}, \bibinfo {author} {\bibfnamefont {I.~J.}\ \bibnamefont {Armyra}},
  \bibinfo {author} {\bibfnamefont {J.}~\bibnamefont {Brillo}}, \bibinfo
  {author} {\bibfnamefont {S.~V.}\ \bibnamefont {Stankus}}, \bibinfo {author}
  {\bibfnamefont {J.}~\bibnamefont {Wu}},\ and\ \bibinfo {author}
  {\bibfnamefont {W.~A.}\ \bibnamefont {Wakeham}},\ }\bibfield  {title}
  {\bibinfo {title} {Reference data for the density and viscosity of liquid
  cadmium, cobalt, gallium, indium, mercury, silicon, thallium, and zinc},\
  }\href@noop {} {\bibfield  {journal} {\bibinfo  {journal} {Journal of
  Physical and Chemical Reference Data}\ }\textbf {\bibinfo {volume} {41}}
  (\bibinfo {year} {2012}{\natexlab{a}})}\BibitemShut {NoStop}%
\bibitem [{\citenamefont {Hardy}(1985)}]{gamma_ga}%
  \BibitemOpen
  \bibfield  {author} {\bibinfo {author} {\bibfnamefont {S.}~\bibnamefont
  {Hardy}},\ }\bibfield  {title} {\bibinfo {title} {The surface tension of
  liquid gallium},\ }\href@noop {} {\bibfield  {journal} {\bibinfo  {journal}
  {Journal of Crystal Growth}\ }\textbf {\bibinfo {volume} {71}},\ \bibinfo
  {pages} {602} (\bibinfo {year} {1985})}\BibitemShut {NoStop}%
\bibitem [{\citenamefont {Assael}\ \emph
  {et~al.}(2012{\natexlab{b}})\citenamefont {Assael}, \citenamefont {Kalyva},
  \citenamefont {Antoniadi}, \citenamefont {Banish}, \citenamefont {Egry},
  \citenamefont {Wu}, \citenamefont {Kaschnitz},\ and\ \citenamefont
  {Wakeham}}]{mu_pb_ag}%
  \BibitemOpen
  \bibfield  {author} {\bibinfo {author} {\bibfnamefont {M.~J.}\ \bibnamefont
  {Assael}}, \bibinfo {author} {\bibfnamefont {A.~E.}\ \bibnamefont {Kalyva}},
  \bibinfo {author} {\bibfnamefont {K.~D.~.}\ \bibnamefont {Antoniadi}},
  \bibinfo {author} {\bibfnamefont {R.~M.}\ \bibnamefont {Banish}}, \bibinfo
  {author} {\bibfnamefont {I.}~\bibnamefont {Egry}}, \bibinfo {author}
  {\bibfnamefont {J.}~\bibnamefont {Wu}}, \bibinfo {author} {\bibfnamefont
  {E.}~\bibnamefont {Kaschnitz}},\ and\ \bibinfo {author} {\bibfnamefont
  {W.~A.}\ \bibnamefont {Wakeham}},\ }\bibfield  {title} {\bibinfo {title}
  {Reference data for the density and viscosity of liquid antimony, bismuth,
  lead, nickel and silver},\ }\href@noop {} {\bibfield  {journal} {\bibinfo
  {journal} {High Temperatures-High Pressures}\ }\textbf {\bibinfo {volume}
  {41}},\ \bibinfo {pages} {161} (\bibinfo {year}
  {2012}{\natexlab{b}})}\BibitemShut {NoStop}%
\bibitem [{\citenamefont {Nogi}\ \emph {et~al.}(1986)\citenamefont {Nogi},
  \citenamefont {Ogino}, \citenamefont {McLean},\ and\ \citenamefont
  {Miller}}]{gamma_pure_metals}%
  \BibitemOpen
  \bibfield  {author} {\bibinfo {author} {\bibfnamefont {K.}~\bibnamefont
  {Nogi}}, \bibinfo {author} {\bibfnamefont {K.}~\bibnamefont {Ogino}},
  \bibinfo {author} {\bibfnamefont {A.}~\bibnamefont {McLean}},\ and\ \bibinfo
  {author} {\bibfnamefont {W.~A.}\ \bibnamefont {Miller}},\ }\bibfield  {title}
  {\bibinfo {title} {The temperature coefficient of the surface tension of pure
  liquid metals},\ }\href@noop {} {\bibfield  {journal} {\bibinfo  {journal}
  {Metallurgical Transactions B}\ }\textbf {\bibinfo {volume} {17B}} (\bibinfo
  {year} {1986})}\BibitemShut {NoStop}%
\bibitem [{\citenamefont {Assael}\ \emph {et~al.}(2010)\citenamefont {Assael},
  \citenamefont {Kalyva}, \citenamefont {Antoniadis}, \citenamefont {Banish},
  \citenamefont {Egry}, \citenamefont {Wu}, \citenamefont {Kaschnitz},\ and\
  \citenamefont {Wakeham}}]{mu_sn}%
  \BibitemOpen
  \bibfield  {author} {\bibinfo {author} {\bibfnamefont {M.~J.}\ \bibnamefont
  {Assael}}, \bibinfo {author} {\bibfnamefont {A.~E.}\ \bibnamefont {Kalyva}},
  \bibinfo {author} {\bibfnamefont {K.~D.}\ \bibnamefont {Antoniadis}},
  \bibinfo {author} {\bibfnamefont {R.~M.}\ \bibnamefont {Banish}}, \bibinfo
  {author} {\bibfnamefont {I.}~\bibnamefont {Egry}}, \bibinfo {author}
  {\bibfnamefont {J.}~\bibnamefont {Wu}}, \bibinfo {author} {\bibfnamefont
  {E.}~\bibnamefont {Kaschnitz}},\ and\ \bibinfo {author} {\bibfnamefont
  {W.~A.}\ \bibnamefont {Wakeham}},\ }\bibfield  {title} {\bibinfo {title}
  {Reference data for the density and viscosity of liquid copper and liquid
  tin},\ }\href@noop {} {\bibfield  {journal} {\bibinfo  {journal} {Journal of
  Physical and Chemical Reference Data}\ }\textbf {\bibinfo {volume} {39}}
  (\bibinfo {year} {2010})}\BibitemShut {NoStop}%
\bibitem [{\citenamefont {Taimatsu}\ and\ \citenamefont
  {Sangiorgi}(1992)}]{gamma_sn}%
  \BibitemOpen
  \bibfield  {author} {\bibinfo {author} {\bibfnamefont {H.}~\bibnamefont
  {Taimatsu}}\ and\ \bibinfo {author} {\bibfnamefont {R.}~\bibnamefont
  {Sangiorgi}},\ }\bibfield  {title} {\bibinfo {title} {Surface tension and
  adsorption in liquid tin-oxygen system},\ }\href@noop {} {\bibfield
  {journal} {\bibinfo  {journal} {Surface Science}\ }\textbf {\bibinfo {volume}
  {261}},\ \bibinfo {pages} {375} (\bibinfo {year} {1992})}\BibitemShut
  {NoStop}%
\bibitem [{\citenamefont {Novakovic}\ \emph {et~al.}(2002)\citenamefont
  {Novakovic}, \citenamefont {Ricci}, \citenamefont {Giuranno},\ and\
  \citenamefont {Gnecco}}]{gamma_pb}%
  \BibitemOpen
  \bibfield  {author} {\bibinfo {author} {\bibfnamefont {R.}~\bibnamefont
  {Novakovic}}, \bibinfo {author} {\bibfnamefont {E.}~\bibnamefont {Ricci}},
  \bibinfo {author} {\bibfnamefont {D.}~\bibnamefont {Giuranno}},\ and\
  \bibinfo {author} {\bibfnamefont {F.}~\bibnamefont {Gnecco}},\ }\bibfield
  {title} {\bibinfo {title} {Surface properties of {Bi?Pb} liquid alloys},\
  }\href@noop {} {\bibfield  {journal} {\bibinfo  {journal} {Surface Science}\
  }\textbf {\bibinfo {volume} {515}},\ \bibinfo {pages} {377} (\bibinfo {year}
  {2002})}\BibitemShut {NoStop}%
\bibitem [{\citenamefont {Williams}\ \emph {et~al.}(2006)\citenamefont
  {Williams}, \citenamefont {Toth},\ and\ \citenamefont {Clarno}}]{mu_flibe}%
  \BibitemOpen
  \bibfield  {author} {\bibinfo {author} {\bibfnamefont {D.}~\bibnamefont
  {Williams}}, \bibinfo {author} {\bibfnamefont {L.}~\bibnamefont {Toth}},\
  and\ \bibinfo {author} {\bibfnamefont {K.}~\bibnamefont {Clarno}},\
  }\href@noop {} {\emph {\bibinfo {title} {Assessment of Candidate Molten Salt
  Coolants for the Advanced High-Temperature Reactor ({AHTR})}}},\ \bibinfo
  {type} {Tech. Rep.}\ \bibinfo {number} {ORNL/TM-2006/12}\ (\bibinfo
  {institution} {Oak Ridge National Laboratory},\ \bibinfo {year}
  {2006})\BibitemShut {NoStop}%
\bibitem [{\citenamefont {Janz}(1988)}]{gamma_mu_flinak}%
  \BibitemOpen
  \bibfield  {author} {\bibinfo {author} {\bibfnamefont {G.~J.}\ \bibnamefont
  {Janz}},\ }\bibfield  {title} {\bibinfo {title} {Thermodynamic and transport
  properties for molten salts: Correlation equations for critically evaluated
  density, surface tension, electrical conductance, and viscosity data},\
  }\href@noop {} {\bibfield  {journal} {\bibinfo  {journal} {Journal of
  Physical and Chemical Reference Data}\ }\textbf {\bibinfo {volume} {17}}
  (\bibinfo {year} {1988})}\BibitemShut {NoStop}%
\bibitem [{\citenamefont {Cantor}\ \emph {et~al.}(1966)\citenamefont {Cantor},
  \citenamefont {Hsu},\ and\ \citenamefont {Ward}}]{flibe_vap_press}%
  \BibitemOpen
  \bibfield  {author} {\bibinfo {author} {\bibfnamefont {S.}~\bibnamefont
  {Cantor}}, \bibinfo {author} {\bibfnamefont {D.~S.}\ \bibnamefont {Hsu}},\
  and\ \bibinfo {author} {\bibfnamefont {W.~T.}\ \bibnamefont {Ward}},\
  }\href@noop {} {\emph {\bibinfo {title} {Vapor Pressures of Fluoride
  Melts}}},\ \bibinfo {type} {Tech. Rep.}\ \bibinfo {number} {ORNL-3913}\
  (\bibinfo  {institution} {Oak Ridge National Laboratory, Reactor Chemistry
  Division},\ \bibinfo {year} {1966})\BibitemShut {NoStop}%
\bibitem [{\citenamefont {Janz}\ \emph {et~al.}(1975)\citenamefont {Janz},
  \citenamefont {Tomkins}, \citenamefont {Allen}, \citenamefont {J.~R.~Downey},
  \citenamefont {Gardner}, \citenamefont {Krebs},\ and\ \citenamefont
  {Singer}}]{rho_flinak}%
  \BibitemOpen
  \bibfield  {author} {\bibinfo {author} {\bibfnamefont {G.~J.}\ \bibnamefont
  {Janz}}, \bibinfo {author} {\bibfnamefont {R.~P.~T.}\ \bibnamefont
  {Tomkins}}, \bibinfo {author} {\bibfnamefont {C.~B.}\ \bibnamefont {Allen}},
  \bibinfo {author} {\bibfnamefont {J.}~\bibnamefont {J.~R.~Downey}}, \bibinfo
  {author} {\bibfnamefont {G.~L.}\ \bibnamefont {Gardner}}, \bibinfo {author}
  {\bibfnamefont {U.}~\bibnamefont {Krebs}},\ and\ \bibinfo {author}
  {\bibfnamefont {S.~K.}\ \bibnamefont {Singer}},\ }\bibfield  {title}
  {\bibinfo {title} {Molten salts: {V}olume 4, part 2, chlorides and mixtures
  electrical conductance, density, viscosity, and surface tension data},\
  }\href@noop {} {\bibfield  {journal} {\bibinfo  {journal} {Journal of
  Physical and Chemical Reference Data}\ }\textbf {\bibinfo {volume} {4}},\
  \bibinfo {pages} {871} (\bibinfo {year} {1975})}\BibitemShut {NoStop}%
\bibitem [{\citenamefont {Cantor}\ \emph {et~al.}(1968)\citenamefont {Cantor},
  \citenamefont {Cooke}, \citenamefont {Dworkin}, \citenamefont {Robbins},
  \citenamefont {Thoma},\ and\ \citenamefont {Watson}}]{fluoride_vap_press}%
  \BibitemOpen
  \bibfield  {author} {\bibinfo {author} {\bibfnamefont {S.}~\bibnamefont
  {Cantor}}, \bibinfo {author} {\bibfnamefont {J.~W.}\ \bibnamefont {Cooke}},
  \bibinfo {author} {\bibfnamefont {A.~S.}\ \bibnamefont {Dworkin}}, \bibinfo
  {author} {\bibfnamefont {G.~D.}\ \bibnamefont {Robbins}}, \bibinfo {author}
  {\bibfnamefont {R.~E.}\ \bibnamefont {Thoma}},\ and\ \bibinfo {author}
  {\bibfnamefont {G.~M.}\ \bibnamefont {Watson}},\ }\href@noop {} {\emph
  {\bibinfo {title} {Physical properties of molten-salt reactor fuel, coolant,
  and flush slats}}},\ \bibinfo {type} {Tech. Rep.}\ \bibinfo {number}
  {ORNL-TM-2316}\ (\bibinfo  {institution} {Oak Ridge National Laboratory,
  Reactor Chemistry Division},\ \bibinfo {year} {1968})\BibitemShut {NoStop}%
\bibitem [{\citenamefont {Liberman}\ \emph {et~al.}(1999)\citenamefont
  {Liberman}, \citenamefont {de~Groot}, \citenamefont {Toor},\ and\
  \citenamefont {Spielman}}]{high_density_z_pinch}%
  \BibitemOpen
  \bibfield  {author} {\bibinfo {author} {\bibfnamefont {M.~A.}\ \bibnamefont
  {Liberman}}, \bibinfo {author} {\bibfnamefont {J.~S.}\ \bibnamefont
  {de~Groot}}, \bibinfo {author} {\bibfnamefont {A.}~\bibnamefont {Toor}},\
  and\ \bibinfo {author} {\bibfnamefont {R.~B.}\ \bibnamefont {Spielman}},\
  }\href@noop {} {\emph {\bibinfo {title} {Physics of High-Density {Z}-Pinch
  Plasmas}}},\ \bibinfo {edition} {1st}\ ed.\ (\bibinfo  {publisher}
  {Springer-Verlag New York},\ \bibinfo {year} {1999})\BibitemShut {NoStop}%
\bibitem [{\citenamefont {Post}\ \emph {et~al.}(1977)\citenamefont {Post},
  \citenamefont {Jensen}, \citenamefont {Kessel}, \citenamefont {Grasberger},\
  and\ \citenamefont {Lokke}}]{coronal_equilibrium}%
  \BibitemOpen
  \bibfield  {author} {\bibinfo {author} {\bibfnamefont {D.~E.}\ \bibnamefont
  {Post}}, \bibinfo {author} {\bibfnamefont {R.~V.}\ \bibnamefont {Jensen}},
  \bibinfo {author} {\bibfnamefont {C.~B.~T.}\ \bibnamefont {Kessel}}, \bibinfo
  {author} {\bibfnamefont {W.~H.}\ \bibnamefont {Grasberger}},\ and\ \bibinfo
  {author} {\bibfnamefont {W.~A.}\ \bibnamefont {Lokke}},\ }\bibfield  {title}
  {\bibinfo {title} {Steady-state radiative cooling rates for low-density,
  high-temperature plasmas},\ }\href@noop {} {\bibfield  {journal} {\bibinfo
  {journal} {Atomic Data and Nuclear Data Tables}\ }\textbf {\bibinfo {volume}
  {20}},\ \bibinfo {pages} {397} (\bibinfo {year} {1977})}\BibitemShut
  {NoStop}%
\bibitem [{\citenamefont {Shearer}(1976)}]{radiation_contraction}%
  \BibitemOpen
  \bibfield  {author} {\bibinfo {author} {\bibfnamefont {J.}~\bibnamefont
  {Shearer}},\ }\bibfield  {title} {\bibinfo {title} {Contraction of {Z}
  pinches actuated by radiation losses},\ }\href@noop {} {\bibfield  {journal}
  {\bibinfo  {journal} {Physics of Fluids}\ }\textbf {\bibinfo {volume} {19}}
  (\bibinfo {year} {1976})}\BibitemShut {NoStop}%
\end{thebibliography}%

\end{document}